\documentclass[aps,preprint,superscriptaddress]{revtex4}
\usepackage{amsfonts}
\usepackage{amsmath}
\usepackage{amssymb}
\usepackage{mathrsfs}
\usepackage{subfigure}
\usepackage{graphicx}
\usepackage{epstopdf}
\usepackage{float}
\usepackage{color}
\usepackage{bm}
\usepackage{ulem}
\usepackage{xcolor}
\usepackage{appendix}
\usepackage{booktabs}
\usepackage{hyperref}
\usepackage{array}
\usepackage{cancel}

\hypersetup{colorlinks=true, linkcolor=blue, filecolor=blue, urlcolor=blue, citecolor=blue}
\begin{document}
\title{Observational Appearances of Magnetically Charged Black Holes in Born-Infeld Electrodynamics}
\preprint{CTP-SCU/2022015}
\author{Shangyu Wen}
\email{shangyuwen@stu.scu.edu.cn}
\affiliation{Center for Theoretical Physics, College of Physics, Sichuan University, Chengdu, 610065, China}
\author{Wei Hong}
\email{weihong@mail.bnu.edu.cn}
\affiliation{Department of Astronomy, Beijing Normal University, Beijing 100875, China}
\author{Jun Tao}
\email{taojun@scu.edu.cn (corresponding author)}
\affiliation{Center for Theoretical Physics, College of Physics, Sichuan University, Chengdu, 610065, China}
\begin{abstract}
    In this paper, we investigate the observational appearances of magnetically charged black holes in Born-Infeld (BI) electrodynamics. We examine the effects of the magnetic charge and the BI parameter on the geodesics with different impact parameters. Using the backward ray tracing method, we investigate how spherically symmetric accretions interact with black hole shadows and photon spheres. The shadows of infalling accretion are darker than that of static ones. Moreover, the radius of the photon sphere is an intrinsic property of the spacetime independent of accretions. We then study how the thin disk models affect the black hole shadows. After obtaining the transfer functions, we divide photons emitted from the thin disk into three categories: direct emission, lens ring, and photon ring. Applying three emission models, we find that the width of the shadow is dominated by the direct emission, the photon ring and the lens ring can hardly be identified by changing the emission models.
\end{abstract}
\maketitle
\section{Introduction}
Recently, the Event Horizon Telescope (EHT) Collaboration processed images of the supermassive black hole (BH) Sgr A* at the center of the galaxy \cite{SgrA_1_2022,SgrA_2_2022,SgrA_3_2022,SgrA_4_2022,SgrA_5_2022,SgrA_6_2022}. Similar to the images of M87* \cite{M87_1_2019,M87_2_2019,M87_3_2019,M87_4_2019,M87_5_2019,M87_6_2019}, these images are relatively dark in the middle and have a bright ring. The central dark region is known as the shadow of the BH, and the bright ring corresponds to the photon sphere \cite{Shoom:2017ril}. The black hole's gravity bends the trajectories of photons emitted by a distant light source behind the black hole, forming a shadow and photon sphere. Modern astrophysical observations suggest that the emissions from BHs are mainly disk-like accretions \cite{Remillard:2006fc,Yuan:2014gma}. The times of photons intersecting with the accretion disk play an important role in the optical appearances of shadows. The photon ring consists of light rays intersecting the accretion disk three or more times \cite{Gralla:2019xty}.

Observing BH shadows is an important way to detect BHs directly, which can enhance our understanding of their properties \cite{PRD_Tsupko_2017,ApJ_Kumar_2020,AAS_Broderick_2022}. Furthermore, the structures of shadows can be used to examine general relativity in strong gravitational fields \cite{NatAstron_Mizuno_2018}. As a result, analyzing the images of BH shadows and photon spheres has become one of the important topics. Schwarzschild BH's shadow was first studied in \cite{MNRAS_Synge_1966}. Sooner, Bardeen et al. studied Kerr BH's shadow \cite{Astrophys_Bardeen_1972}. Through analytical calculation, the first figures of BH shadow with a rotating accretion disk are proposed in \cite{Astronomy_Luminet_1979}. In some modified gravity theories, the shadows of rotating BHs were also investigated \cite{Dastan:2016vhb,Long:2020wqj}. Compared to the EHT observations for M87* and SgrA*, there have been studies scrutinizing visible shapes of these black holes for the brightest point in accretion disk \cite{Dokuchaev:2019pcx, Dokuchaev:2020wqk}. Physical origin of the dark spot at the SgrA* image was studied \cite{Dokuchaev:2022aku}. For invisible black holes, their silhouettes was also investigated \cite{Dokuchaev:2019jqq}. The research on BH shadows has been expanded to many aspects in recent years \cite{Qiao:2022jlu,Chakhchi:2022fls,Lin:2022ksb,Belhaj:2022kek,Sun:2022wya,Guo:2021bhr,Zhong:2021mty,Cunha:2016wzk,CPC_He_2022,Gan:2021xdl,Shaikh:2021yux,Heydari-Fard:2022jdu,Bambi:2019tjh,Vagnozzi:2019apd,Vagnozzi:2020quf,Roy:2021uye,Chen:2022nbb,Vagnozzi:2022moj}.

There are some pieces of evidence suggesting that magnetic fields may exist near BHs. After two years of in-depth research and data processing, the EHT Collaboration unveiled the shadow of M87* with polarized light \cite{M87_7_2021,M87_8_2021}. The image clearly showed that the shadow of the BH is affected by magnetic fields, implying there probably exists a strong magnetic field surrounding the BH. Just recently, the motion of the hot spot orbiting Sgr A* is predicted to be relevant to the magnetic fields surrounding the BH \cite{Wielgus:2022heh}. In recent decades, there have been many theoretical studies on magnetic black hole properties, like formation, evolution and classical instability \cite{Lee_PRL_1992,Lee_PRD_1992,Lee_PRL_1994}.

To eliminate the singularity of point-like charge in Maxwell electrodynamics, Born and Infeld proposed the BI electrodynamics \cite{Nature_BI_1933}, inspired by special relativity and modified the Lagrangian of electromagnetic fields. In addition to BI electrodynamics, many other theories have been proposed to modify Maxwell electrodynamics (MED) to avoid singularity \cite{PRB_Kruglov_2012,PRD_Kruglov_2007,IJGMMP_Kruglov_2015,Kruglov:2020aqm}. These Maxwell's theory extensions are usually known as nonlinear electrodynamics (NLED). Among NLEDs, BI is distinguished for it is the only one that ensures no birefringence \cite{AIP_Kerner_2001}. Surprisingly, BI Lagrangian can be exactly derived from low energy string theory \cite{PLB_Fradkin_1985}. In addition, BI has been studied as coupling with general relativity \cite{PR_Hoffmann_1935}. In recent years, Einstein-Born-Infeld gravity has been studied in various papers \cite{Babar:2021nst,Wang:2020ohb,Jafarzade:2020ova,Ali:2022zox,Yang:2021bhv,Zhang:2021kha,Falciano:2021kdu,Hendi:2018cfr,Mazharimousavi:2014vza,Dehghani:2019noj,Jing:2020sdf,Gan:2019jac,Liang:2019dni,Tao:2017fsy,Bi:2020vcg}. Magnetically charged BH has been studied with the presence of BI NLED \cite{Kruglov:2017mpj}, as well as other NLEDs \cite{Kruglov:2020aqm,Kruglov:2019okd}. Considering two NLED models and comparing the shadows of these models to the observation of M87*, an upper bound on the magnetic charge of BHs was given \cite{Allahyari:2019jqz}. The potential astrophysical signatures of magnetically charged BHs are also studied \cite{Ghosh:2020tdu}. According to \cite{BenavidesGallego:2018odl}, the magnetic charge strongly affects the outer horizons and the ergospheres of rotating BHs, but the circular photon orbits are not explicitly dependent on magnetic charge. Furthermore, the effects of strong magnetic fields on shadows are examined \cite{PRD_Haroldo_2021}. Theoretically, photons move along the null geodesics of the spacetime governed by the BH. Being affected by NLED, photons move along the null geodesics of the effective metric rather than the background metric \cite{IJMPD_Bergliaffa_2004,Kruglov:2020tes}. Besides, it was shown that magnetically charged BH can have regular solution \cite{Bronnikov:2000vy} when NLED becomes Maxwell electrodynamics at weak fields. And a good example is the rational NLED model first introduced by Kruglov \cite{Kruglov:rationalNLED}, which has not only limited electric field but also the possibility to regard the mass of the electron as pure electromagnetic energy. After that, BH shadows' radii with the presence of rational NLED are investigated, which results are compatible with M87* data \cite{Kruglov:2020tes}. We can compare the basic properties and observational appearance of BI BHs with magnetic regular BHs in rational NLED.

This paper investigates the observational appearances of magnetically charged BI BH. The shadows and photon spheres of BHs are derived using the backward ray tracing method \cite{Astronomy_Luminet_1979}. Additionally, we investigate the impact of the accretion's dynamics and shapes on the shadows. This paper is organized as follows. In section \ref{geo}, we investigate the metric and the trajectories of photons deflected by a magnetically charged BI BH. Additionally, we discuss the range of the magnetic charge and the BI parameter. Section \ref{spheric} studies the shadows and photon spheres with spherical accretion. In section \ref{ring}, the shadows produced by thin accretion disks are investigated. In section \ref{concl}, we discuss and conclude our works. Moreover, we compares the basic properties and shadows of BI BHs with magnetic regular BHs in rational NLED in Appendix \ref{appendixComparison}.

\section{Geodesics\label{geo}}
\subsection{The metric}
In this section, we derive the metric of BI BHs with magnetic charge in static spherically symmetric spacetime. The Einstein-Born-Infeld action is defined as follows \cite{PRD_Cai_2004}
\begin{align}
    \mathcal{S}=\int\sqrt{-g}\left[ R+\mathcal{L}(F) \right]{\rm{d}}^{4}x,
\end{align}
where $R$ is the Ricci scalar, $F=F_{\mu\nu}F^{\mu\nu}$. The BI Lagrangian $\mathcal{L}(F)$ takes the form as
\begin{align}
    \mathcal{L}(F)=4\beta^{2}\left( 1-\sqrt{1+\frac{F}{2\beta^{2}}} \right),
    \label{LF}
\end{align}
where $\beta$ is called the BI parameter with the dimension of mass. When $\beta\to\infty$, $\mathcal{L}(F)$ degenerates into the standard electromagnetic form. Varying the action with the gauge field $A_{\mu}$, the BI equation yields
\begin{align}
    \partial_{\mu}\left(\frac{\sqrt{-g}F^{\mu\nu}}{\sqrt{1+\frac{F}{2\beta^2}}}\right)=0.
\end{align}
Combining with the Einstein equation, the energy momentum tensor takes the form as
\begin{align}
    T_{\mu\nu}=g_{\mu\nu}\mathcal{L}(F)+\frac{4F_{\mu\alpha}F_{\nu}^{\alpha}}{\sqrt{1+\frac{F}{2\beta^2}}}.
    \label{EngMomTensor}
\end{align}

Restricting to static and spherically symmetric spacetime, one can have the ansatz
\begin{align}
    {\rm{d}}s^{2}=-f(r){\rm{d}}t^{2}+\frac{1}{f(r)}{\rm{d}}r^{2}+r^{2}({\rm{d}}\theta^{2}+\sin^{2}\theta{\rm{d}}\phi^{2}),
    \label{ds^2_raw}
\end{align}
and the electromagnetic tensor generated by magnetically charged BH takes the form as
\begin{align}
    F_{\theta\phi}=-F_{\phi\theta}=P\sin\theta,
    \label{EMfield}
\end{align}
where $P$ is a parameter representing the magnetic charge.

The metric function $f(r)$ for magnetically charged BI BHs is given in \cite{EPJC_He_2022}
\begin{align}
    f(r)=1-\frac{2M}{r}+\frac{2\beta^{2}r^{2}}{3}-\frac{2\beta^{2}r^{2}}{3}{_{2}F_{1}}\left(-\frac{3}{4},-\frac{1}{2};\frac{1}{4};-\frac{P^{2}}{r^{4}\beta^{2}}\right),
    \label{fr}
\end{align}
where $M$ is the mass of the BH, and $_{2}F_{1}$ is the hypergeometric function. When $\beta\to+\infty$, $f(r)$ degenerates into the form of magnetically charged Reissner-Nordström (RN) BH.
\begin{align}
    f(r)\overset{\text{expand at }\beta=+\infty}{=\!=\!=\!=\!=\!=\!=\!=\!=}1-\frac{2M}{r}+\frac{P^{2}}{r^{2}}+O\left(\frac{1}{\beta}\right)^{2}.
\end{align}

\begin{figure}[htb]
    \centering
    \includegraphics[width=9cm]{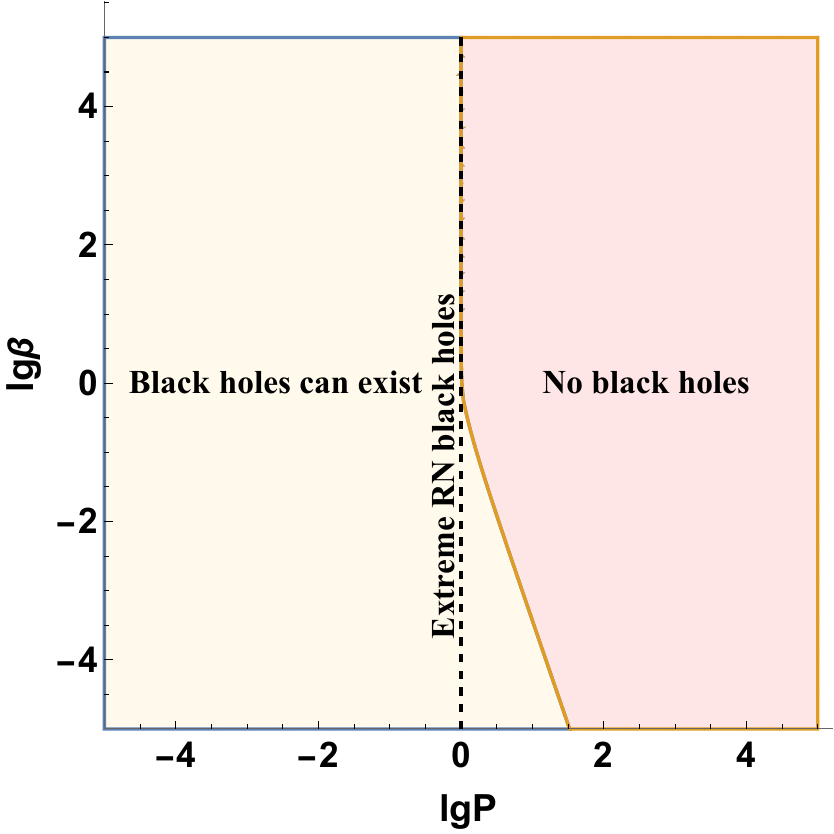}
    \caption{Region plot in $\lg\beta-\lg P$ space for magnetically charged BI BH with fixed $M=1$. Black holes can only exist in the light yellow area. The critical value of the magnetic charge $P$ increases with the decrease of $\beta$. The black dashed line ($P=1$) represents extreme RN BHs. To display a wider plot range, we use logarithmic coordinates.}
    \label{lgbeta_lgP}
\end{figure}

In addition, the equation $f(r)=0$ has two roots. The larger one $r_{\text{h}}$ represents the radius of the event horizon. Since it is not easy to obtain analytical solutions of $r_{\text{h}}$, its numerical solutions of different $\beta$ are listed in Table \ref{NumSol_changeBeta}. Setting the minimum of $f(r)$ to zero will yield the critical value of $P$ with respect to different $\beta$, as Fig. \ref{lgbeta_lgP} and Table \ref{NumSol_Pcrit} show.

\begin{table}[!htbp]
    \setlength{\tabcolsep}{5mm}
    \begin{center}
        \begin{tabular}[b]{c|cccccc}
            \midrule[2pt]
            $\beta$ & $0.05$ & $0.1$ & $0.15$ & $0.2$ & $0.3$ & $\infty$ \\
            \toprule[1pt]
            $P_{\text{crit}}$ & 2.35678 & 1.87057 & 1.63410 & 1.48467 & 1.29698 & 1.00000 \\
            \midrule[2pt]
        \end{tabular}
    \end{center}
    \centering
    \caption{Data of critical value of magnetic charge $P_{\text{crit}}$ for different $\beta$ with fixed $M=1$. The critical value of magnetic charge decreases as the BI effect weakens until $\beta=\infty$, when $P_{\text{crit}}$ becomes the same as magnetically charged RN BH.}
    \label{NumSol_Pcrit}
\end{table}

\subsection{The innermost stationary circular orbit\label{isco}}
Next, we derive the innermost stationary circular orbit (ISCO) for massive particles. The Lagrangian of a particle with unit mass, $\mathcal{L}_{m}$, has the form as
\begin{align}
    \mathcal{L}_{m}=\frac{1}{2}g_{\mu\nu}\dot{x}^{\mu}\dot{x}^{\nu},
    \label{Lagrangian_massive}
\end{align}
where the dot on $x^{\mu}$ represents the derivative with respect to the proper time $\tau$. For massive particles, there are two conserved quantities: energy $E_{m}$ and angular momentum $L_{m}$, which are obtained by
\begin{align}
    E_{m}=-\frac{\partial {\mathcal{L}}_{m}}{\partial \dot{t}}=f(r)\dot{t},\qquad L_{m}=\frac{\partial \mathcal{L}_{m}}{\partial \dot{\phi}}=r^{2}\dot{\phi}.
    \label{conservations}
\end{align}
Substituting Eq. (\ref{conservations}) into Eq. (\ref{Lagrangian_massive}), noting that $\mathcal{L}_{m}=-1$, we derive the equation of radial motion
\begin{align}
    \dot{r}^{2}+\frac{L_{m}^2}{r^{2}}f(r)+f(r)=E_{m}^{2}.
    \label{rdot_massive}
\end{align}
We define the effective potential of massive particles to be
\begin{align}
    V_{m}(r)=\frac{L_{m}^2}{r^{2}}f(r)+f(r).
    \label{eff_potential_massive}
\end{align}
The radius of the ISCO satisfies
\begin{align}
    \left.\frac{{\rm{d}}V_{m}}{{\rm{d}}r}\right|_{r=r_{\text{ISCO}}}=0,\qquad \left.\frac{{\rm{d}}^{2}V_{m}}{{\rm{d}}r^{2}}\right|_{r=r_{\text{ISCO}}}>0.
    \label{ISCO_conditions}
\end{align}
Solving Eq. (\ref{ISCO_conditions}) numerically, we obtain the radii of the innermost stationary circular orbit $r_{\text{ISCO}}$ for different BI parameters $\beta$ and magnetic charges $P$ in Tables \ref{NumSol_changeBeta} and \ref{NumSol_changeP}. 

\begin{figure}[htb]
    \centering
    \includegraphics[width=10cm]{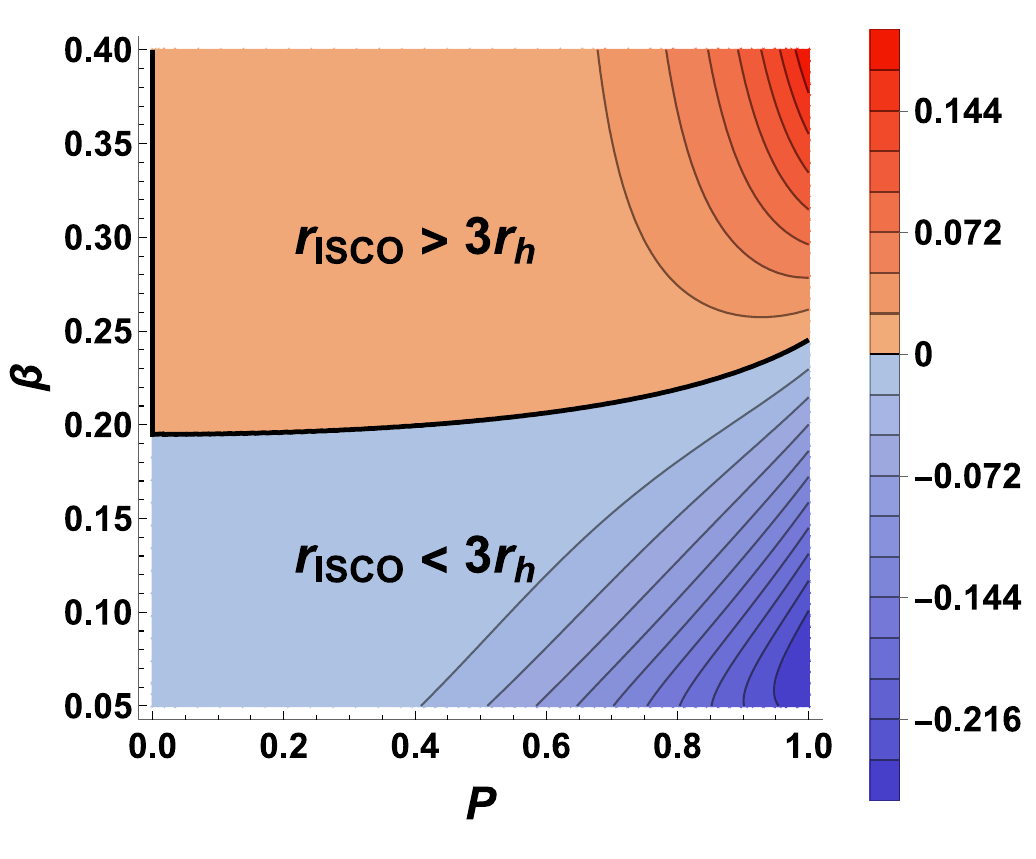}
    \caption{Contour plot in $\beta-P$ space with fixed $M=1$. The colors represent the value of $r_{\text{ISCO}}-3r_{\text{h}}$, and the contours are the isosurfaces of $r_{\text{ISCO}}-3r_{\text{h}}$. On the bolded curve, $r_{\text{ISCO}}=3r_{\text{h}}$. The impact of the magnetic charge $P$ is opposite to the BI parameter $\beta$ below the bolded curve, and vice versa.}
    \label{risco3rh}
\end{figure}

We calculate the values of $r_{\text{ISCO}}-3r_{\text{h}}$ for different parameters $\beta$ and $P$ in Tables \ref{NumSol_changeBeta} and \ref{NumSol_changeP} to compare the spacetime of magnetically charged BI BH to that of Schwarzschild BH, for $r_{\text{ISCO}}=3r_{\text{h}}$ in Schwarzschild spacetime. The non-monotonicity of $r_{\text{ISCO}}-3r_{\text{h}}$ shown in Tables \ref{NumSol_changeBeta} and \ref{NumSol_changeP} implies that there is a critical value of $\beta$ that determines at which $r_{\text{ISCO}}-3r_{\text{h}}$ switches from negative to positive. Fig. \ref{risco3rh} is the contour plot of $r_{\text{ISCO}}-3r_{\text{h}}$ in $\beta$-$P$ space. We start from $\beta=0.05$ rather than $\beta=0$ since numerical instability occurs at $\beta$ close to 0. Changes in $\beta$ and $P$ have contrary impacts on $r_{\text{ISCO}}-3r_{\text{h}}$ on each side of the critical values. In the red region, $r_{\text{ISCO}}-3r_{\text{h}}$ increases with $P$ and $\beta$. In the blue region, the impact of $P$ on $r_{\text{ISCO}}-3r_{\text{h}}$ is opposite to that of $\beta$. In other words, the impact of $P$ is contrary to that of $\beta$ if it is less than the critical value and vice versa. When $P=0$, a magnetically charged BI BH degenerates into a Schwarzschild BH, regardless of the BI effect, as the leftmost bolded line in Fig. \ref{risco3rh} shows.

\subsection{The trajectories of photons}

In observations, for black hole shadows in nonlinear electrodynamics, photons dispersion relations need to be revised due to NLED effects. From a different perspective, the dispersion relations can be interpreted as that of the null vector with the geometry described by the effective metric. 
It means that the motion of lights can be viewed as electromagnetic waves propagating through a classical dispersive medium in a nontrivial vacuum \cite{Dittrich:1998fy}. The effective metric for magnetic BI black holes takes the form as \cite{Plebansky:1968, Kim:2022xum, EPJC_He_2022}
\begin{align}
    {\rm{d}} s_{\text{eff}}^{2}=k(r)\left[-f(r){\rm{d}} t^{2}+\frac{1}{f(r)}{\rm{d}} r^{2}+h(r)({\rm{d}} \theta^{2}+\sin^{2}\theta {\rm{d}}\phi^{2})\right],
    \label{eff_ds^2}
\end{align}
where
\begin{align}
    k(r)=\sqrt{1+\frac{P^{2}}{\beta^{2}r^{4}}},\quad h(r)=r^{2}\left(1+\frac{P^{2}}{\beta^{2}r^{4}}\right).
\end{align}
The BI Parameter $\beta$ characterizes the strength of NLED effects. In the limit $\beta\rightarrow \infty$, both the metric function in Eq. (\ref{fr}) and the effective metric function in Eq. (\ref{eff_ds^2}) reduces to the ordinary one for the RN black hole. Therefore, for magnetic black holes if one uses ordinary metric without NLED effects, the observation appearance reduces to the one for magnetic RN black holes, which has been extensively studied \cite{Grenzebach:2014fha, Jusufi:2020agr, Kumaran:2022soh, Mandal:2022oma}.

Thus, we can derive the trajectories of photons. The Lagrangian of a photon is
\begin{align}
    \mathcal{L}=\frac{1}{2}G_{\mu\nu}\dot{x}^{\mu}\dot{x}^{\nu},
    \label{Lagrangian}
\end{align}
where $G_{\mu\nu}$ are the covariant components of the effective metric, and the dot on $x^{\mu}$ represents the derivative with respect to the affine parameter $\lambda$. We simply focus on the equatorial plane ($\theta=\pi/2$ and $\dot{\theta}=0$) because the metric is spherically symmetric. Moreover, since the Lagrangian does not explicitly depend on coordinates $t$ and $\phi$, there are two conserved quantities along the geodesics
\begin{align}
    E=-\frac{\partial\mathcal{L}}{\partial\dot{t}}=k(r)f(r)\dot{t},\qquad L=\frac{\partial\mathcal{L}}{\partial\dot{\phi}}=k(r)h(r)\dot{\phi}.
    \label{conservations_photons}
\end{align}
For photons $\mathcal{L}=0$, one can derive the following equations representing the motion of photons
\begin{align}
    -\frac{E^{2}}{k(r)f(r)}+\frac{k(r)}{f(r)}\dot{r}^{2}+\frac{L^{2}}{k(r)h(r)}=0.
    \label{rdot1}
\end{align}
The equations of photons' motion are derived by redefining the affine parameter $\lambda$ into $\lambda/L$ and setting the impact parameter $b=L/E$.
\begin{align}
    \dot{t}=\frac{1}{bk(r)f(r)},
    \label{tdot2}
\end{align}
\begin{align}
    \dot{\phi}=\pm \frac{1}{k(r)h(r)},
    \label{phidot2}
\end{align}
\begin{align}
    k^{2}(r)\dot{r}^{2}=\frac{1}{b^{2}}-\frac{f(r)}{h(r)},
    \label{rdot2}
\end{align}
where $\pm$ in Eq. (\ref{phidot2}) represents the counterclockwise and clockwise motion of the photon respectively. We define the effective potential $V(r)=f(r)/h(r)$. Specially, when $\dot{r}|_{r=r_{\text{ph}}}=0$ and ${\ddot{r}}|_{r=r_{\text{ph}}}=0$, it suggests that the photon can circle the BH for infinite times in an unstable orbit called photon sphere, which indicates
\begin{align}
    V(r_{\text{ph}})=\frac{1}{b_{\text{ph}}^{2}},\quad \left.\frac{{\rm{d}}V}{{\rm{d}}r}\right|_{r=r_{\text{ph}}}=0,
    \label{eff_potential}
\end{align}
where $r_{\text{ph}}$ is the radius of the photon sphere, $b_{\text{ph}}$ is the corresponding impact parameter. Since it is not easy to derive an analytical expression of $r_{\text{ph}}$ and $b_{\text{ph}}$, we list their numerical solutions in Tables \ref{NumSol_changeBeta} and \ref{NumSol_changeP}.

As Table \ref{NumSol_changeBeta} shows, $r_{\text{h}}$, $r_{\text{ISCO}}$, $r_{\text{ph}}$, and $b_{\text{ph}}$ decrease monotonically when $\beta$ increases. Despite that $r_{\text{ph}}$ drops steeper than $r_{\text{h}}$ as $\beta$ increases, $r_{\text{h}}<r_{\text{ph}}$ is always satisfied because a BI BH degenerates into an RN BH when $\beta\to\infty$.

\begin{table}[htbp]
    \setlength{\tabcolsep}{3mm}
    \begin{center}
        \begin{tabular}[b]{c|cccccc}
            \midrule[2pt]
            $\beta$ & $0.05$ & $0.1$ & $0.15$ & $0.2$ & $0.3$ & $\infty$ \\
            \toprule[1pt]
            $r_{\text{h}}$ & 1.88896 & 1.87594 & 1.87134 & 1.86928 & 1.86758  & 1.86603 \\
            $r_{\text{ISCO}}$ & 5.62174 & 5.61066 & 5.60845 & 5.60766 & 5.60710 & 5.60664 \\
            $(r_{\text{ISCO}}-3r_{\text{h}})\times 10^{3}$ & -45.1400 & -17.1501 & -5.57996 & -0.17694 & 4.36839 & 8.56722 \\
            $r_{\text{ph}}$ & 4.02971 & 3.52690 & 3.09423 & 3.06541 & 2.94078 & 2.82288 \\
            $b_{\text{ph}}$ & 6.57031 & 5.86395 & 5.31786 & 5.27579 & 5.11771 & 4.96791 \\
            \midrule[2pt]
        \end{tabular}
    \end{center}
    \centering
    \caption{Data of $r_{\text{h}}$, $r_{\text{ISCO}}$, $r_{\text{ph}}$, and $b_{\text{ph}}$ for different $\beta$ with fixed $M=1$, $P=0.5$.}
    \label{NumSol_changeBeta}
\end{table}

Numerical solutions of different $P$ are listed in Table \ref{NumSol_changeP}. With the increase of $P$, $r_{\text{h}}$, and $r_{\text{ISCO}}$ decrease, while $r_{\text{ph}}$ and $b_{\text{ph}}$ increase monotonically.

\begin{table}[htbp]
    \setlength{\tabcolsep}{3mm}
    \begin{center}
        \begin{tabular}[b]{c|cccccc}
            \midrule[2pt]
            $P$ & 0.3 & 0.4 & 0.5 & 0.6 & 0.7 & 0.9 \\
            \toprule[1pt]
            $r_{\text{h}}$ & 1.95519 & 1.92050 & 1.87594 & 1.82139 & 1.75661 & 1.59483 \\
            $r_{\text{ISCO}}$ & 5.86321 & 5.75424 & 5.61066 & 5.42955 & 5.20713 & 4.61999 \\
            $(r_{\text{ISCO}}-3r_{\text{h}})\times 10^{3}$ & -2.37168 & -7.25105 & -17.1501 & -34.6012 & -62.6931 & -164.491 \\
            $r_{\text{ph}}$ & 3.23266 & 3.37470 & 3.52690 & 3.68206 & 3.83617 & 4.13392 \\
            $b_{\text{ph}}$ & 5.49393 & 5.67308 & 5.86395 & 6.05783 & 6.24991 & 6.61985 \\
            \midrule[2pt]
        \end{tabular}
    \end{center}
    \centering
    \caption{Data of $r_{\text{h}}$, $r_{\text{ISCO}}$, $r_{\text{ph}}$, and $b_{\text{ph}}$ for different $P$ with fixed $M=1$, $\beta=0.1$.}
    \label{NumSol_changeP}
\end{table}

\begin{figure}[htb]
    \begin{center}
        \subfigure[$\ V(r)$]{\includegraphics[width=6cm]{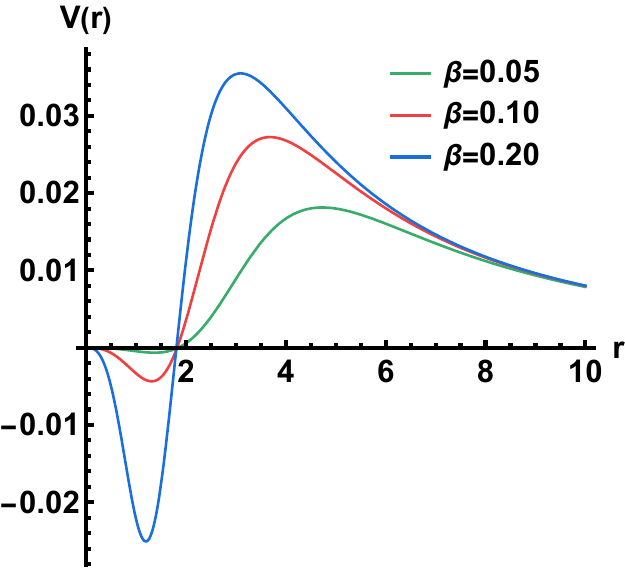}\label{geo_change_beta_V(r)}}
        \subfigure[$\ \beta=0.05$]{\includegraphics[width=6cm]{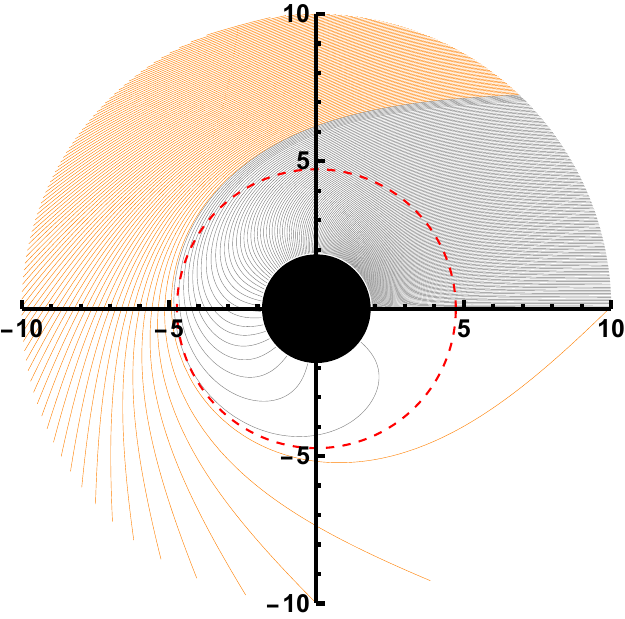}\label{geo_change_beta_beta=0.05}}
        \subfigure[$\ \beta=0.1$]{\includegraphics[width=6cm]{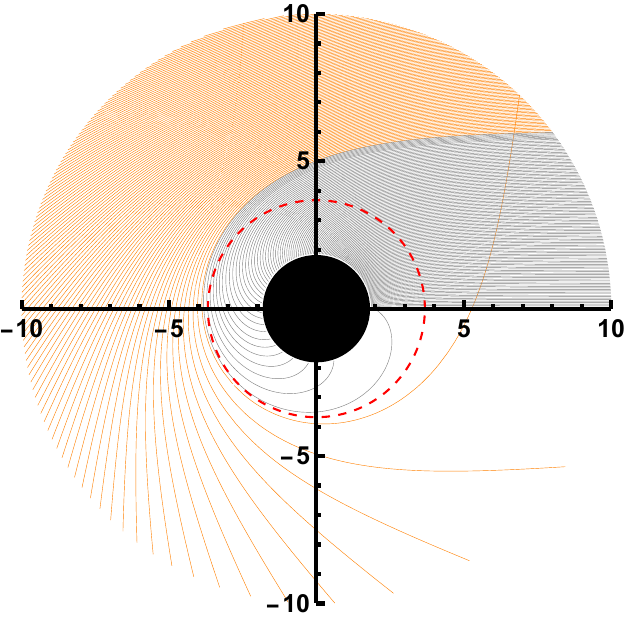}\label{geo_change_beta_beta=0.1}}
        \subfigure[$\ \beta=0.2$]{\includegraphics[width=6cm]{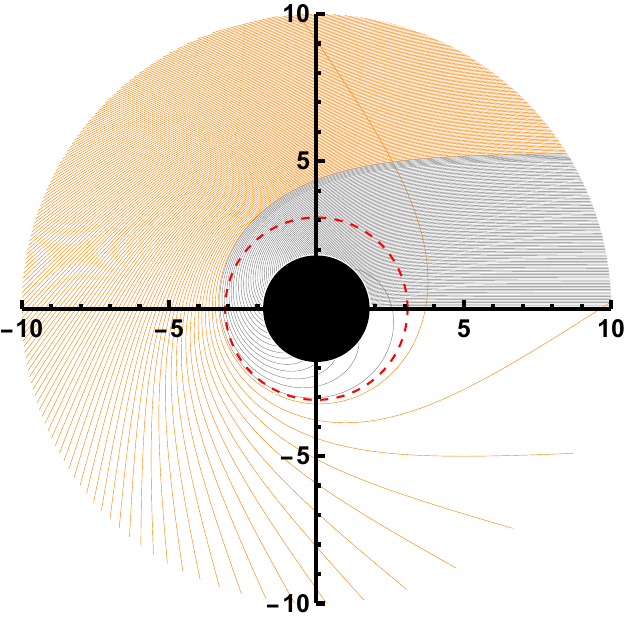}\label{geo_change_beta_beta=0.2}}
    \end{center}
    \caption{The effective potential $V(r)$ for different BI parameter $\beta$ are shown in the top-left panel. The other figures are trajectories of photons for different values of $\beta$ with fixed $P=0.6$ and $M=1$.}
    \label{geo_change_beta}
\end{figure}

\begin{figure}[htb]
    \begin{center}
        \subfigure[$\ V(r)$]{\includegraphics[width=6cm]{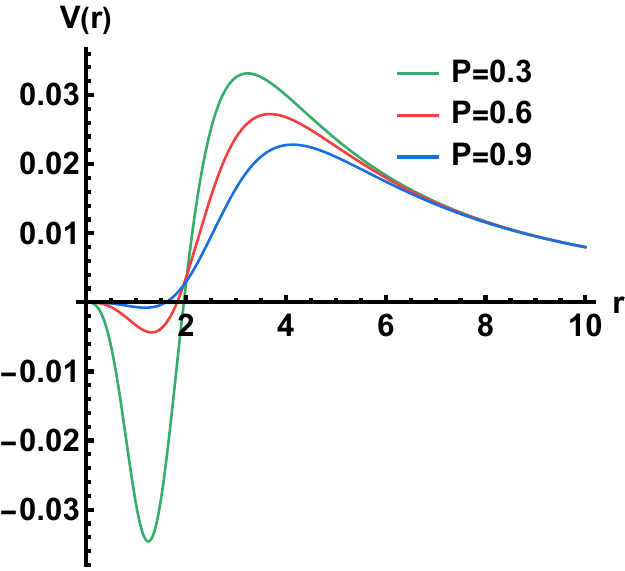}\label{geo_change_P_V(r)}}
        \subfigure[$\ P=0.3$]{\includegraphics[width=6cm]{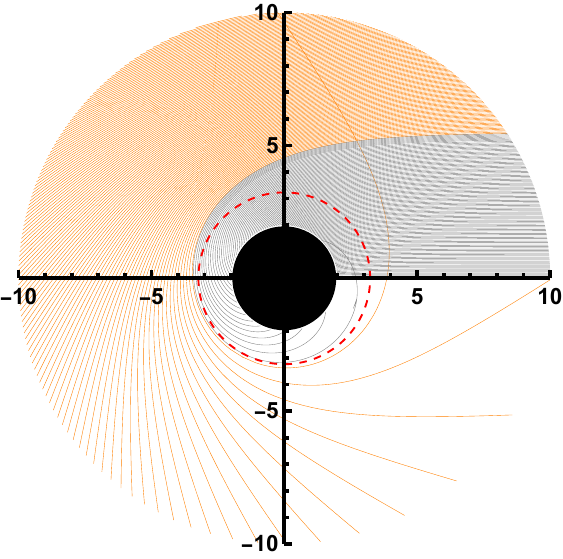}\label{geo_change_P_P=0.3}}
        \subfigure[$\ P=0.6$]{\includegraphics[width=6cm]{geoP0.6beta0.1.pdf}\label{geo_change_P_P=0.6}}
        \subfigure[$\ P=0.9$]{\includegraphics[width=6cm]{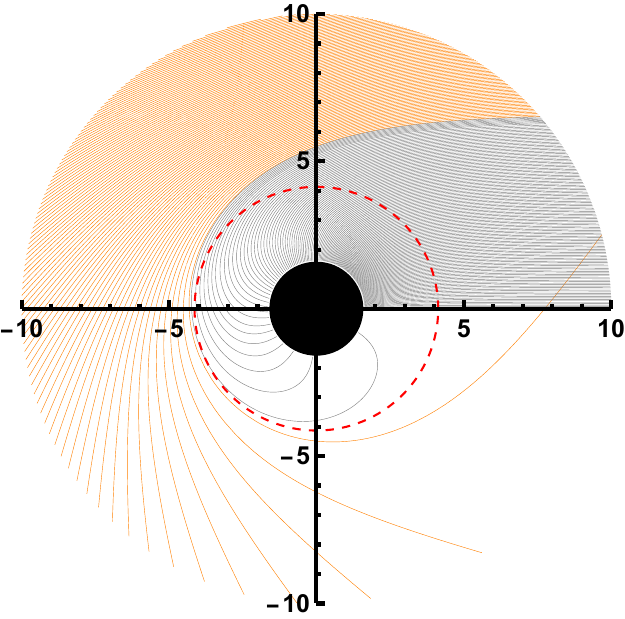}\label{geo_change_P_P=0.9}}
    \end{center}
    \caption{The effective potential $V(r)$ for different magnetic charges $P$ are shown in the top-left panel. The other figures are trajectories of photons for different values of $P$ with fixed $\beta=0.1$ and $M=1$.}
    \label{geo_change_P}
\end{figure}

The effective potentials with respect to different parameters $\beta$ and $P$ are shown respectively in Figs. \ref{geo_change_beta_V(r)} and \ref{geo_change_P_V(r)}. The trajectories of photons can be derived by combining Eqs. (\ref{phidot2}) and (\ref{rdot2}), yielding

\begin{align}
    \frac{{\rm{d}}r}{{\rm{d}}\phi}=\pm h(r)\sqrt{\frac{1}{b^{2}}-\frac{f(r)}{h(r)}}\equiv F(r),
    \label{geodesics}
\end{align}
where $\pm$ represents the photons moving radially inward and outward, respectively. We can obtain the null geodesics by integrating Eq. (\ref{geodesics}). For $b>b_{\text{ph}}$, photons approach the BH before reaching the closest approach $r=r_{0}$, then escape towards infinity. The closest approach $r_{0}$ can be obtained by solving $F(r)=0$. In the backward ray tracing method \cite{Astronomy_Luminet_1979}, $r_{0}$ is significant for determining the photons' trajectories as it can be the lower integral limit. Meanwhile, for $b<b_{\text{ph}}$, photons approach the BH continuously before falling into it.

We set the photons incident from infinity on the x-axis. Firstly, we study the geodesics for fixed $P=0.6$ and $M=1$ as $\beta$ varies. The numerical solutions of $r_{\text{h}}$, $r_{\text{ph}}$, and $b_{\text{ph}}$ are shown in Table \ref{NumSol_changeBeta}. Interestingly, $r_{\text{ph}}$ and $b_{\text{ph}}$ decrease dramatically with the increase of $\beta$ when it is less than 0.2. When $\beta$ increases, $r_{\text{ph}}$ and $b_{\text{ph}}$ slowly decline and gradually converge to the case of RN BH. Therefore, we select $\beta=0.05,\ 0.1,\ 0.2$ to investigate the photons' trajectories in Fig. \ref{geo_change_beta}.

In Figs. \ref{geo_change_beta_beta=0.05}, \ref{geo_change_beta_beta=0.1}, and \ref{geo_change_beta_beta=0.2}, the black disk represents the BH. And the orange and grey trajectories represent photons originating with $b>b_{\text{ph}}$ and $b<b_{\text{ph}}$, respectively. In the case of photons coming from $b=b_{\text{ph}}$, they continually approach the BH and keep rotating around it in an unstable orbit located at $r=r_{\text{ph}}$, i.e., the photon sphere, which is shown by red dashed lines. The BI parameter significantly affects the radius of the photon sphere and the curvature of the geodesics. However, as $\beta$ increases, the decrease in $r_{\text{h}}$ becomes so insignificant that it is scarcely visible in those figures.

Then, we study the influence of magnetic charge on geodesics with fixed $\beta=0.1$ and $M=1$. Using the same method, we list the numerical solutions of $r_{\text{h}}$, $r_{\text{ph}}$, and $b_{\text{ph}}$ in Table \ref{NumSol_changeP}. The geodesics of different magnetic charges are plotted in Fig. \ref{geo_change_P_P=0.3}, \ref{geo_change_P_P=0.6}, and \ref{geo_change_P_P=0.9}. As the BH gets more magnetically charged or the BI effect becomes stronger, the range between the photon spheres and the event horizon extends.

\section{Shadows and photon spheres with spherical accretions\label{spheric}}
\subsection{Stationary spherical accretions\label{static}}
In this subsection, we study the BH shadows of static and spherically symmetric accretions. We employ the backward ray tracing method to study the specific intensity received by a distant observer \cite{Astronomy_Luminet_1979}. The emissivity of photons can be expressed as \cite{CPC_He_2022}
\begin{align}
    j(\nu_{\text{e}})\propto\rho(r)P(\nu_{\text{e}}),
    \label{j}
\end{align}
where $\rho(r)$ is the density of photons at a given radius $r$, and $P(\nu_{\text{e}})$ is the probability of photons emitting at a given frequency $\nu_{\text{e}}$. For spherical accretions, we assume that the density of photons follows a logarithmic normal distribution \cite{Zhang:2022osx}
\begin{align}
    \rho(r)=\frac{1}{r}\sqrt{\frac{\gamma}{\pi}}{\rm{e}}^{-\gamma\ln^{2}\frac{r}{r_{\text{m}}}},
    \label{rho}
\end{align}
where $\gamma$ is a constant that affects how fast the density of photons decays with respect to the distance from the BH, $r_{\text{m}}$ is the median of the logarithmic normal distribution. Since the photon density reaches the maximum at the photon sphere, we select $r=r_{\text{ph}}$ as the highest value of the probability density of the lognormal distribution. Solving $\left.\frac{{\rm{d}}\rho(r)}{{\rm{d}}r}\right|_{r=r_{\text{ph}}}=0$, one can obtain 
\begin{align}
    r_{\text{m}}=r_{\text{ph}}{\rm{e}}^{\frac{1}{2\gamma}}.
    \label{rph_and_rm}
\end{align}

We introduce a non-monochromatic light source satisfying a normal distribution with central frequency $\nu_{\text{c}}$, which takes the form as
\begin{align}
    f(\nu)=\frac{1}{\sigma\sqrt{2\pi}}\exp\left(-\frac{(\nu-\nu_{\text{c}})^{2}}{2\sigma^{2}}\right).
\end{align}
In this paper, we only take the frequency within $\nu_{\text{c}}-\sqrt{2}\sigma<\nu<\nu_{\text{c}}+\sqrt{2}\sigma$ into account and ignore the rest. The probability of a photon whose frequency falls within the aforementioned range is
\begin{align}
    P(\nu_{\text{c}})=\int_{\nu_{\text{c}}-\sqrt{2}\sigma}^{\nu_{\text{c}}+\sqrt{2}\sigma}f(\nu){\rm{d}}\nu\approx 0.84270.
    \label{probability}
\end{align}

\begin{figure}[htb]
    \centering
    \includegraphics[width=9cm]{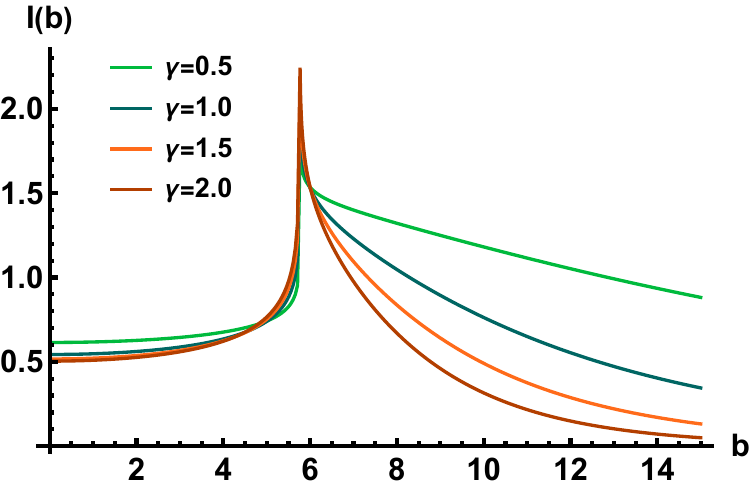}
    \caption{The specific intensity with different $\gamma$ in static spherical accretion with fixed $\beta=0.1,\ P=0.6$, and $M=1$.}
    \label{Ib_static_change_gamma}
\end{figure}

Combining Eqs. (\ref{j}), (\ref{rho}), (\ref{rph_and_rm}), and (\ref{probability}) together, one can derive
\begin{align}
    j(\nu_{\text{e}})\propto \frac{1}{r}\sqrt{\frac{\gamma}{\pi}}{\rm{e}}^{-\gamma\ln^{2}\frac{r}{r_{\text{m}}}}.
    \label{j_int}
\end{align}
Integrating Eq. (\ref{j_int}) along the geodesics and setting the coefficient of proportion to be unit, we obtain the specific intensity
\begin{align}
    I(b)=\int_{\text{geo}} g^{3} j(\nu_{\text{e}}){\rm{d}}l_{\text{prop}},
    \label{Ib}
\end{align}
where $g=\nu_{\text{o}}/\nu_{e}$ is the redshift factor, $\nu_{\text{o}}$ is the frequency received by the observer, ${\rm{d}}l_{\text{prop}}$ is the infinitesimal proper length observed in the instantaneous rest frame of the photon. The redshift factor $g$ is given by
\begin{align}
    g=\frac{p_{\alpha}u_{\text{o}}^{\alpha}}{p_{\beta}u_{\text{e}}^{\beta}},
    \label{g}
\end{align}
where $p_{\alpha}$ and $p_{\beta}$ are the 4-momenta of the photons received and those emitted, respectively. And $u_{\text{o}}^{\alpha}$ is the 4-velocity of the observer, $u_{\text{e}}^{\beta}$ is the 4-velocity of the static accretions. Setting the observer at $r=\infty$, from $G_{\mu\nu}u_{\text{e}}^{\mu}u_{\text{e}}^{\nu}=1$ we derive
\begin{align}
    \begin{split}
        u_{\text{o}}^{\alpha}&=\left(1,\ 0,\ 0,\ 0\right),\\
        u_{\text{e}}^{\beta}&=\left( \sqrt{\frac{1}{k(r)f(r)}},\ 0,\ 0,\ 0 \right).
    \end{split}
    \label{4velocity_static}
\end{align}
Substituting $p_{t}=-E$ into the Hamiltonian of photons $\mathcal{H}=G^{\mu\nu}p_{\mu}p_{\nu}=0$, and redefining the affine parameter $\lambda$ to $\lambda/L$, we derive
\begin{align}
    p_{\mu}=\left( -1,\ \pm\sqrt{\frac{1}{f(r)}\left( \frac{1}{f(r)}-\frac{b^2}{h(r)} \right)},\ 0,\ \pm b \right),
    \label{4momentum_static}
\end{align}
where the first positive and negative sign, respectively, represent photons moving radially inward and outward. And $\pm b$ represents photons moving counterclockwise and clockwise. Substituting Eqs. (\ref{4velocity_static}) and (\ref{4momentum_static}) into Eq. (\ref{g}), we get
\begin{align}
    g=\sqrt{k(r)f(r)}.
    \label{g_static_derived}
\end{align}
And the infinitesimal proper length can be obtained as follows
\begin{align}
    {\rm{d}}l_{\text{prop}}=\pm \sqrt{G_{ij}{\rm{d}}x^{i}{\rm{d}}x^{j}}=\pm \sqrt{\frac{k(r)}{f(r)}+h(r)k(r)\left(\frac{{\rm{d}}\phi}{{\rm{d}}r}\right)^{2}}{\rm{d}}r,
    \label{dl}
\end{align}
where ${\rm{d}}\phi/{\rm{d}}r=1/F(r)$, can be obtained from Eq. (\ref{geodesics}). Substituting Eqs. (\ref{j_int}) and (\ref{dl}) into Eq. (\ref{Ib}), we obtain the specific intensity $I$ of the impact parameter $b$. For different decay factors $\gamma$, we plot the specific intensity in Fig. \ref{Ib_static_change_gamma}.

From Fig. \ref{Ib_static_change_gamma} we can see that the specific intensity grows slightly as $b<b_{\text{ph}}$. But then surges to the maximum as $b$ approaches $b_{\text{ph}}$, steadily decline when $b>b_{\text{ph}}$, and gradually approaches 0 until $b=\infty$. What is of importance in this figure is that the maximum of the specific intensity located at $b=b_{\text{ph}}$, as a result of the photons revolving around the BH multiple times. Therefore, $b_{\text{ph}}$ is an intrinsic property of the effective metric independent of $\gamma$.

\begin{figure}[htb]
    \begin{center}
        \subfigure[$\ I(b)$]{\includegraphics[width=6cm]{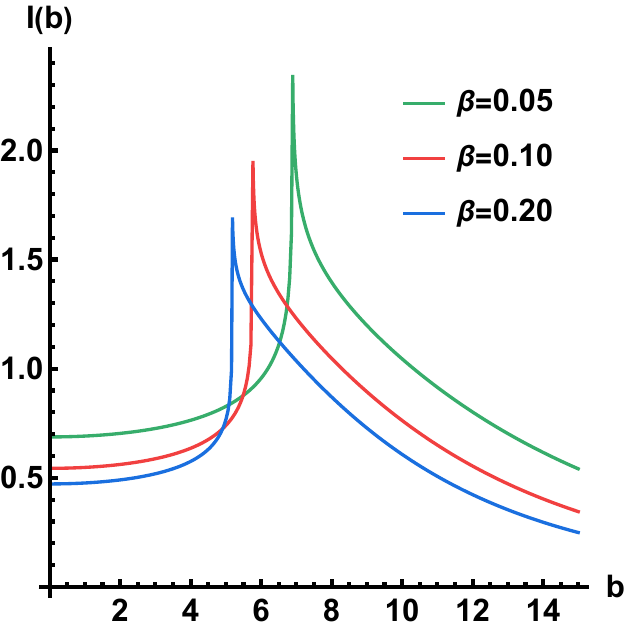}\label{I(b)_static_change_beta}}
        \subfigure[$\ \beta=0.05$]{\includegraphics[width=6cm]{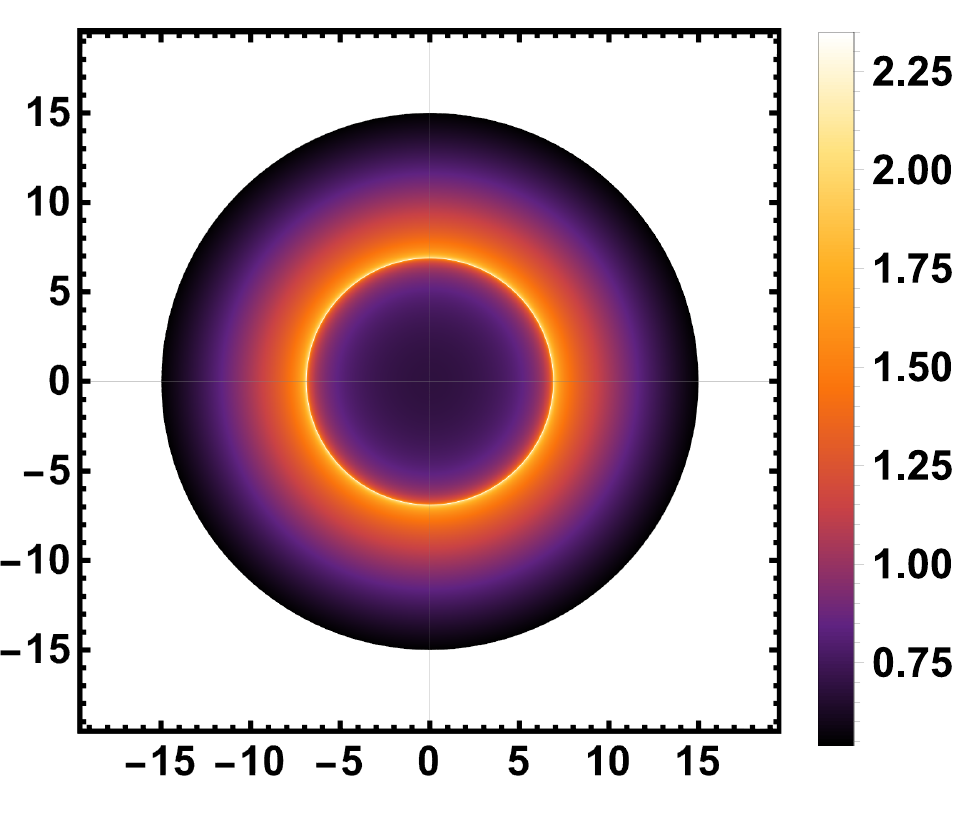}\label{shadow_static_change_beta_beta=0.05}}
        \subfigure[$\ \beta=0.1$]{\includegraphics[width=6cm]{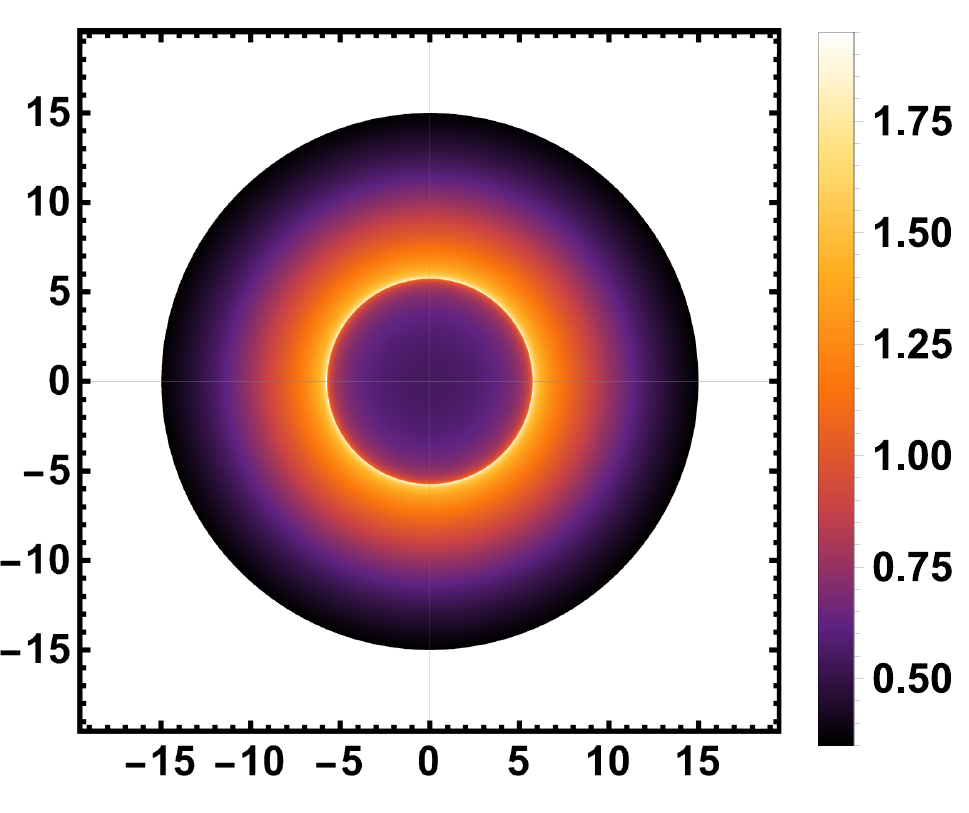}\label{shadow_static_change_beta_beta=0.1}}
        \subfigure[$\ \beta=0.2$]{\includegraphics[width=6cm]{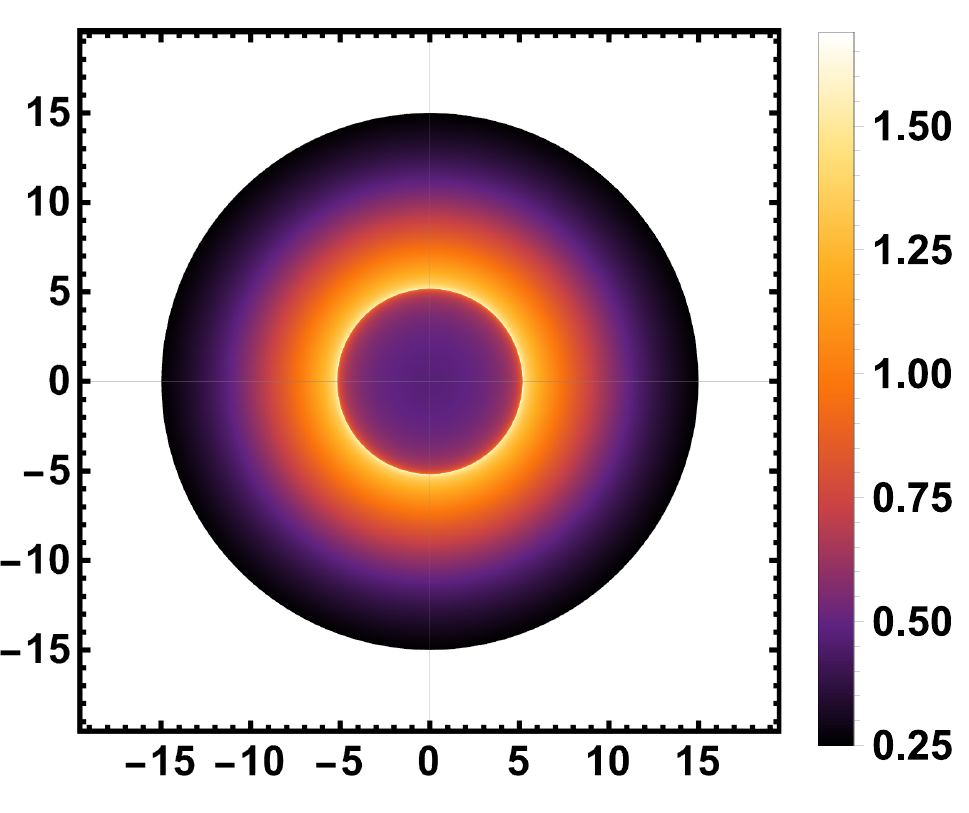}\label{shadow_static_change_beta_beta=0.2}}
    \end{center}
    \caption{The specific intensity $I$ with respect to different impact parameters $b$ in the top-left panel. And the other figures are the BH shadows with static accretions for different BI parameters $\beta$ with $P=0.6$ and $M=1$.}
    \label{shadows_static_change_beta}
\end{figure}

\begin{figure}[htb]
    \begin{center}
        \subfigure[$\ I(b)$]{\includegraphics[width=6cm]{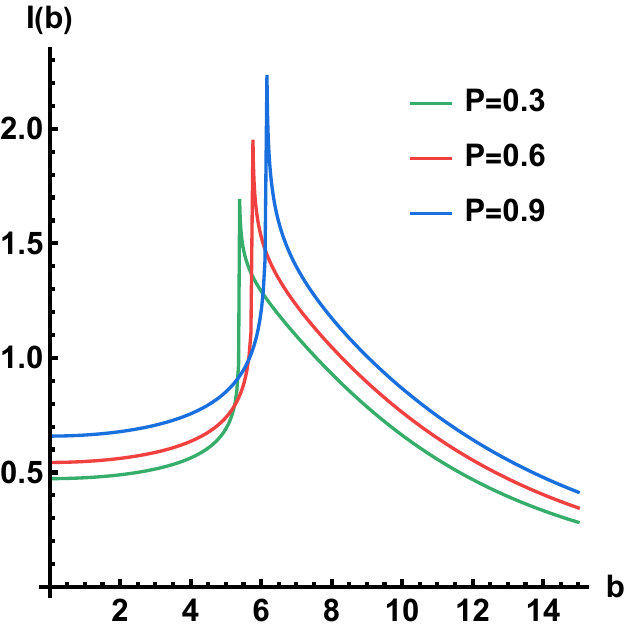}\label{I(b)_static_change_P}}
        \subfigure[$\ P=0.3$]{\includegraphics[width=6cm]{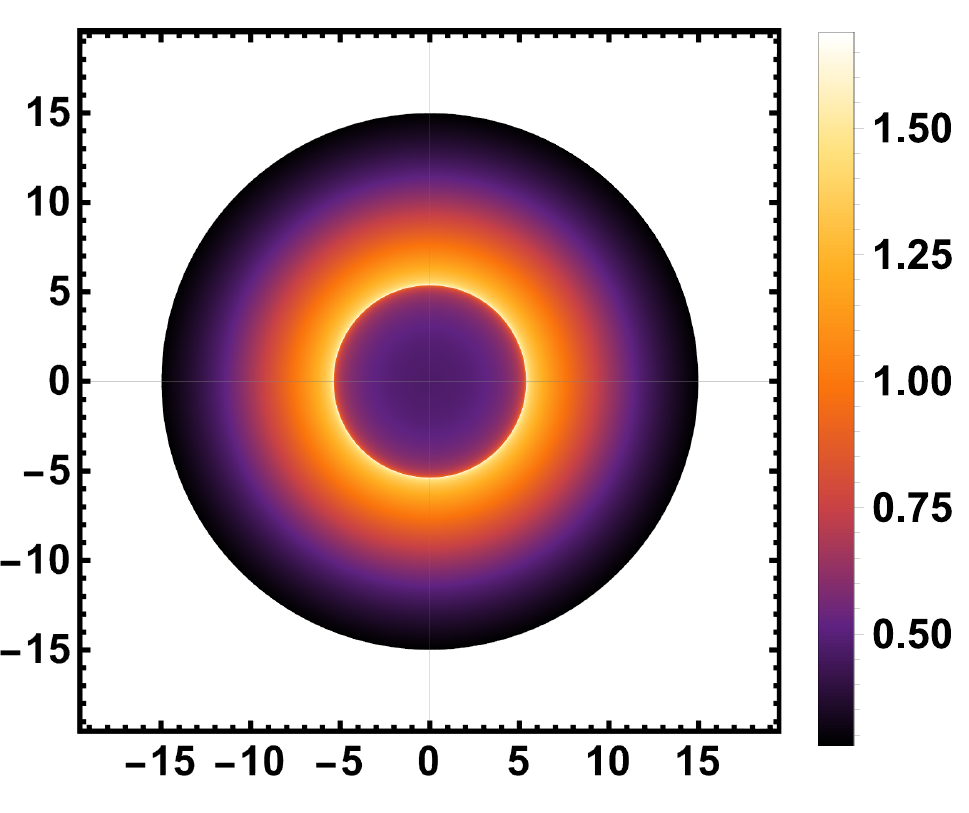}\label{shadow_static_change_P_P=0.3}}
        \subfigure[$\ P=0.6$]{\includegraphics[width=6cm]{shadow_static_P0.6beta0.1.pdf}\label{shadow_static_change_P_P=0.6}}
        \subfigure[$\ P=0.9$]{\includegraphics[width=6cm]{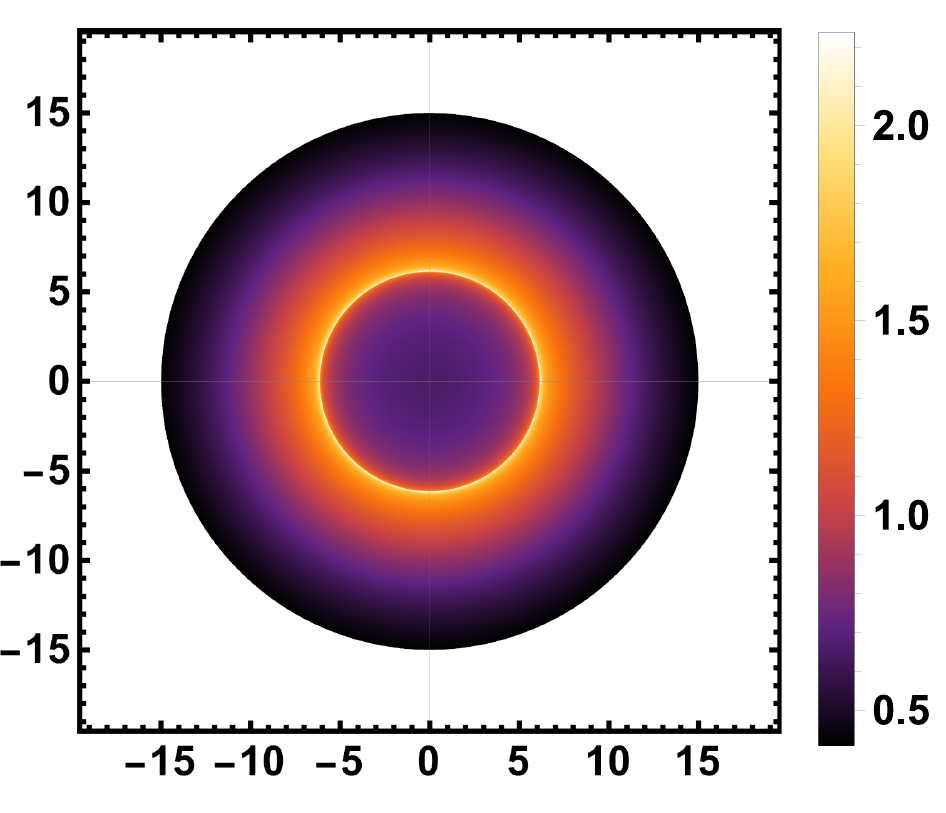}\label{shadow_static_change_P_P=0.9}}
    \end{center}
    \caption{The specific intensity $I$ with respect to different impact parameters $b$ in the top-left panel. And the other figures are the BH shadows with static accretions for different magnetic charge with $\beta=0.1$ and $M=1$.}
    \label{shadows_static_change_P}
\end{figure}

For convenience, we set $\gamma=1$ in the following discussions. We set the inclination angle of the observer to be zero due to the spherically symmetric of the spacetime. The impact parameter $b$ determines the light ray's trajectory and thus determines the observed intensity \cite{EPJC_Zeng_2020}. Thereon, we can obtain the specific intensity $I(b)$ from Eq. (\ref{Ib}) and then plot the simulated shadows in Figs. \ref{shadows_static_change_beta} and \ref{shadows_static_change_P}. The specific intensity is represented by several colors, which are brighter if the specific intensity is greater, and vice versa.

The specific intensity and the shadows with static accretions when $P=0.6$ and $M=1$ with respect to different $\beta$ are shown in Fig. \ref{shadows_static_change_beta}. The photon spheres are the brightest rings in Figs. \ref{shadow_static_change_beta_beta=0.05}, \ref{shadow_static_change_beta_beta=0.1}, and \ref{shadow_static_change_beta_beta=0.2}. And the radii of the photon spheres decrease continuously with the increase of $\beta$, but the decreasing speed gradually slows down, which is consistent with Table \ref{NumSol_changeBeta}. Also, we discuss the specific intensities and shadows in the case of $\beta=0.1,\ M=1$ as $P$ varies in Fig. \ref{shadows_static_change_P} by applying the same method above. The radii of the photon spheres obviously increase as the magnetic charge increases, as described in Table. \ref{NumSol_changeP}. In summary, the average radial distance between the photons received by the observer and the BH would increase, causing a rise in the specific intensity, if the BH gets more magnetically charged or the effect of BI becomes more significant.

Additionally, We summarized and compared our results for magnetic BI BH with magnetic regular BH in rational NLED \cite{Kruglov:rationalNLED} (Appendix \ref{appendixComparison}). Observational appearances of regular BH in rational NLED were also investigated with different magnetic charge and rational NLED parameter $\beta_r$ . We infer that BI and the rational NLED model have similar effects on the observational appearances of magnetic black holes with static spherical accretions.

\subsection{Infalling spherical accretions\label{infalling}}
In this subsection, we study the shadows of free-falling spherically symmetric accretions in a static and spherically symmetric spacetime. The accretions have radial velocity pointing to the BH in this model. Similarly, we also employ Eq. (\ref{Ib}) to describe the intensities of the shadows. In infalling spherical accretions, the redshift factor $g$ can be calculated through Eq. (\ref{g}). The 4-velocity of a spherically symmetric infalling accretion influenced by the magnetic charged BI BH is \cite{NPB_ZENG_2022}
\begin{align}
    u_{\text{e}}^{\beta}=\left( \frac{1}{A(r)},\ -\sqrt{\frac{1-A(r)}{A(r)B(r)}},\ 0,\ 0 \right).
    \label{4velocity_infalling}
\end{align}
For simplicity, we set $A(r)=k(r)f(r)$ and $B(r)=1/[k(r)f(r)]$. The observer's position is fixed as the same as that in section \ref{static}, through the same method in obtaining Eqs. (\ref{4velocity_static}) and (\ref{4momentum_static}), we derive
\begin{align}
    g=\frac{1}{\frac{1}{k(r)f(r)}\pm\sqrt{1-k(r)f(r)}\sqrt{\frac{1}{f(r)}\left(\frac{1}{f(r)}-\frac{b^{2}}{h(r)}\right)}},
    \label{g_infalling_derived}        
\end{align}
\begin{figure}[htb]
    \begin{center}
        \subfigure[$\ \beta=0.05.$]{
            \includegraphics[width=6cm]{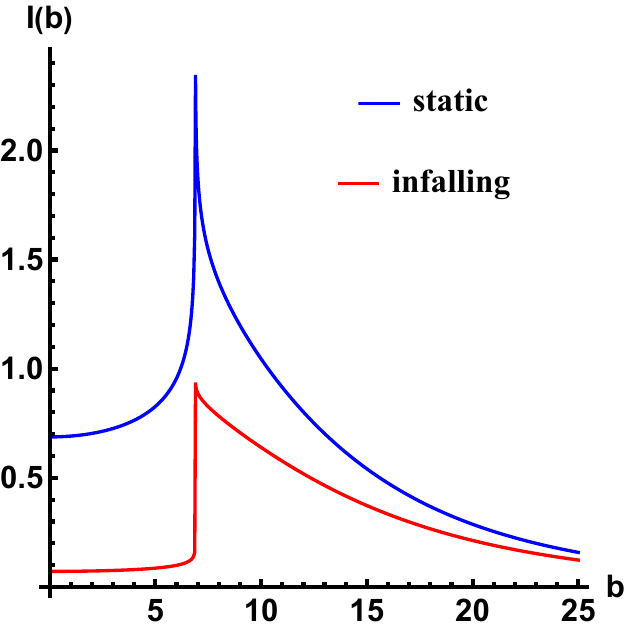}\label{I(b)_staticinfalling_change_beta_beta=0.05}
            \hspace{2mm}
            \includegraphics[width=6cm]{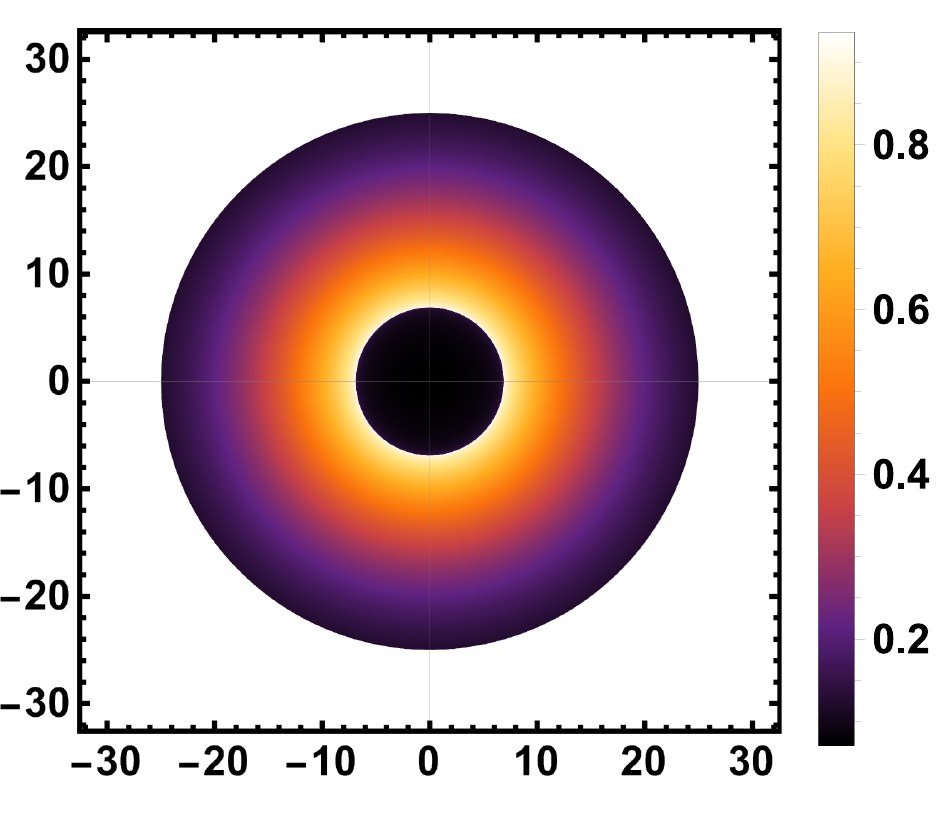}\label{shadow_infalling_change_beta_beta=0.05}
        }
        \subfigure[$\ \beta=0.2.$]{
            \includegraphics[width=6cm]{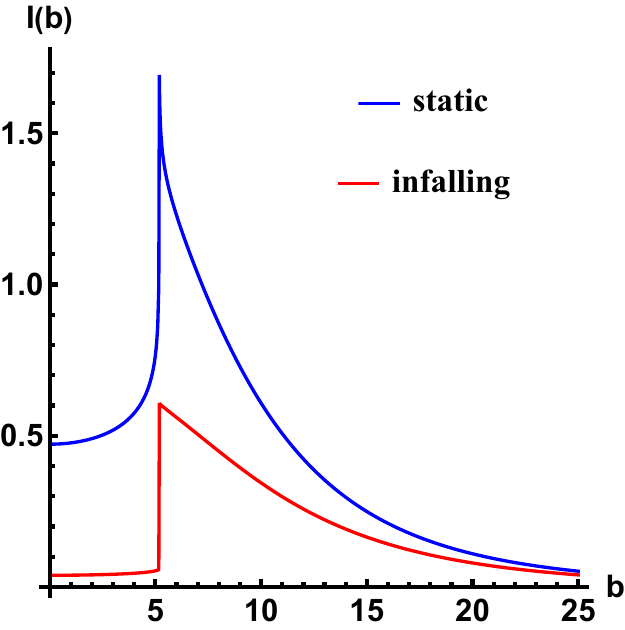}\label{I(b)_staticinfalling_change_beta_beta=0.2}
            \hspace{2mm}
            \includegraphics[width=6cm]{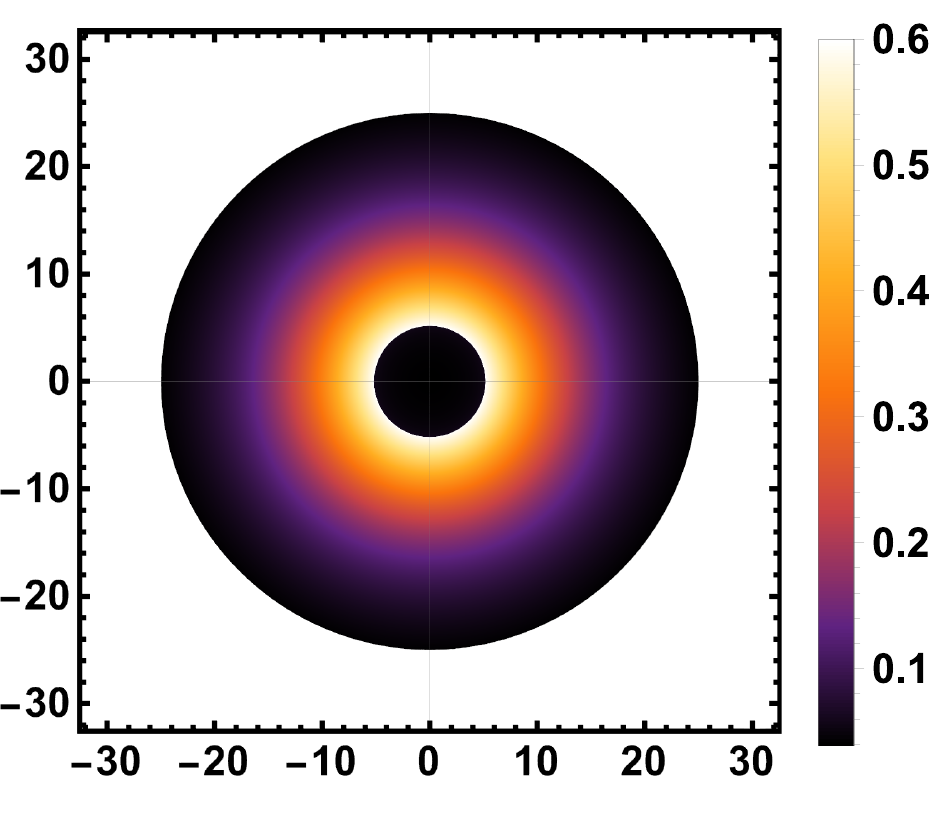}\label{shadow_infalling_change_beta_beta=0.2}
        }
    \end{center}
    \caption{The specific intensity $I$ with respect to different impact parameters $b$ in the left panels with static and infalling accretions for different $\beta$ with fixed $P=0.6$ and $M=1$. And the shadows with these parameters for infalling accretions are shown in the right panels.}
    \label{staticinfalling_change_beta}
\end{figure}
where the $\pm$ sign represents the photons moving radially inward and outward respectively. Therefore, we obtain the specific intensity $I$ with the impact parameter $b$ in Fig. \ref{staticinfalling_change_beta} by combining Eqs. (\ref{Ib}) and (\ref{g_infalling_derived}). The specific intensity in infalling accretion decreases slower compared with the static one. We extend the plot range to $b\leq 25$ to display the properties of shadows more comprehensively.
\begin{figure}[htb]
    \begin{center}
        \subfigure[$\ P=0.3.$]{
            \includegraphics[width=6cm]{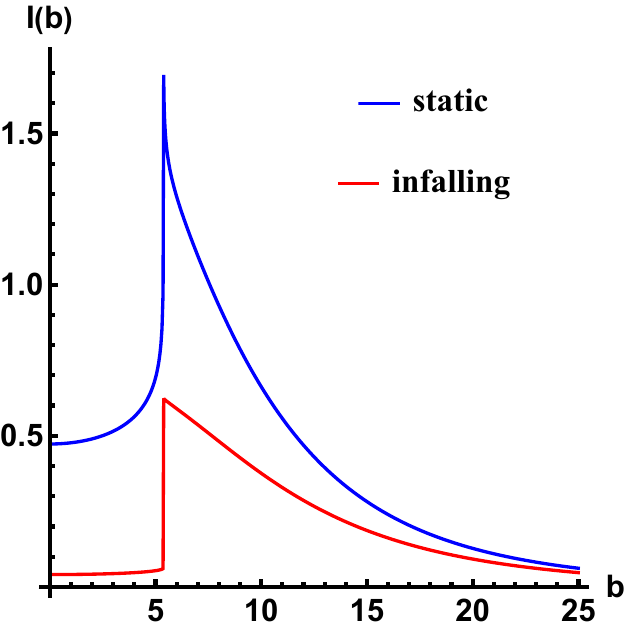}\label{I(b)_staticinfalling_change_P_P=0.3}
            \hspace{2mm}
            \includegraphics[width=6cm]{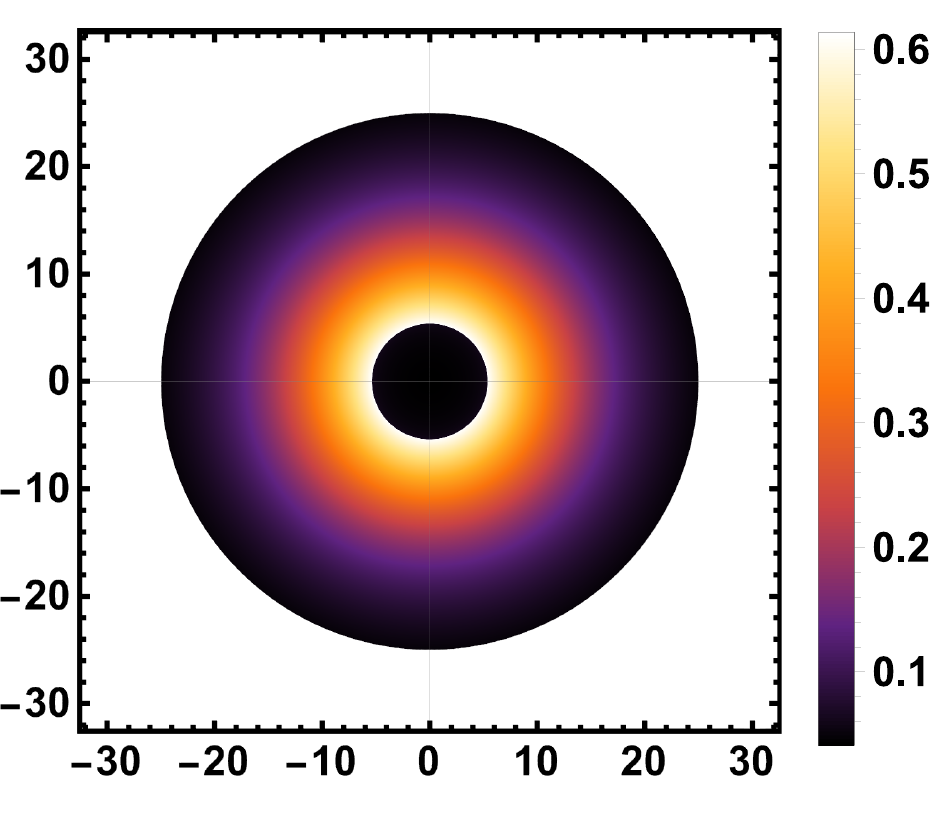}\label{shadow_infalling_change_P_P=0.3}
        }
        \subfigure[$\ P=0.9.$]{
            \includegraphics[width=6cm]{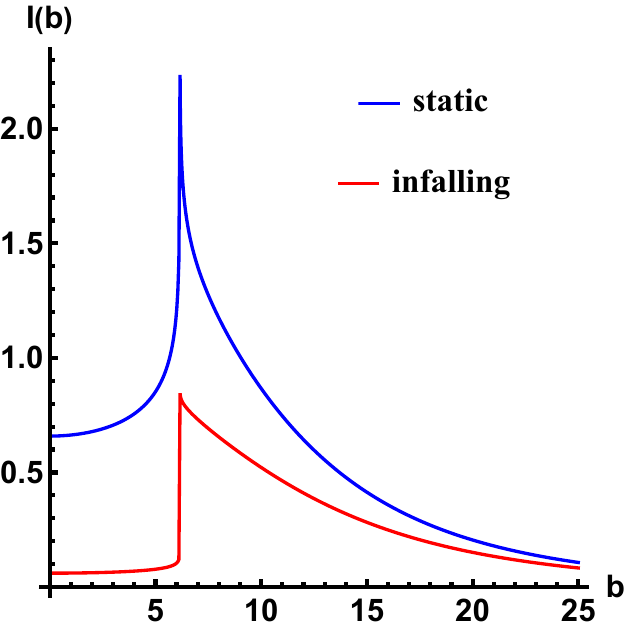}\label{I(b)_staticinfalling_change_P_P=0.9}
            \hspace{2mm}
            \includegraphics[width=6cm]{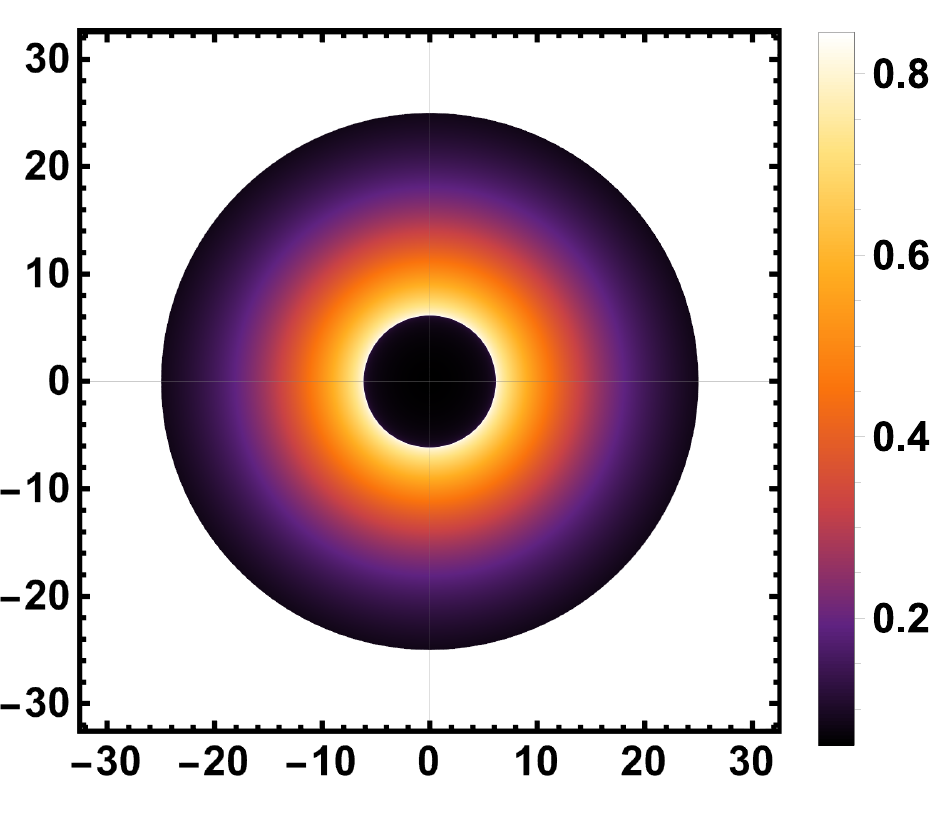}\label{shadow_infalling_change_P_P=0.9}
        }
    \end{center}
    \caption{The specific intensity $I$ with respect to different impact parameters $b$ in the left panels with static and infalling accretions for different $P$ with fixed $\beta=0.1$ and $M=1$. And the shadows with these parameters for infalling accretions are shown in the right panels.}
    \label{staticinfalling_change_P}
\end{figure}

As shown in Fig. \ref{staticinfalling_change_beta}, the specific intensities of infalling accretion are obviously darker than those of the static one. Compared to static accretion, the infalling one has a smaller redshift factor $g$. As a result, the impact of redshift on infalling accretion is more obvious than that on static accretion. If the photons' impact parameter $b$ is small, their trajectories will get close to the BH. As a consequence, the integration along the geodesics contributes less to the specific intensity, resulting in a significantly lower value. Interestingly, the specific intensity of infalling accretion decreases inapparently as $b\to 0$ when $b<b_{\text{ph}}$, compared to that of static accretion. That is because the average radial distance between photons and the BH increases as the impact parameter increases, and the impact of the redshift factor on photons attenuates with their distance from the BH.

Additionally, the specific intensity of either static or infalling accretion, converges to zero as $b$ approaches infinity. Firstly, the average distance between photons and the BH positively correlates to the impact parameter $b$. And the photon density decreases with the distance from the BH. As a result, the density of photons is less when the impact parameter is greater. Secondly, as illustrated in the end of the previous paragraph, the effect of the redshift factor on photons becomes weaker with increasing distance from the BH.

It is also noteworthy that, in both static and infalling accretion, the impact parameter corresponding to the peak of the specific intensity is always the same. This is because, in this paper, we ignore the other factors that may affect the photons' equivalent metric, such as the electromagnetic interaction between the accretion matters and photons \cite{Zhang:2022osx}.

We also investigate the impact of different magnetic charges on the specific intensity with infalling spherical accretion with fixed $\beta=0.1$ in Fig. \ref{staticinfalling_change_P}. Due to the slow increase of $b_{\text{ph}}$ as the BH gets more magnetically charged (see Table \ref{NumSol_changeP}), the differences between Figs. \ref{shadow_infalling_change_P_P=0.3} and \ref{shadow_infalling_change_P_P=0.9} are not as obvious as that between Figs. \ref{shadow_infalling_change_beta_beta=0.05} and \ref{shadow_infalling_change_beta_beta=0.2}. The BH shadow in the observational appearance shrinks as the BH becomes more magnetically charged or the influence of BI becomes more significant. The reasons are the same as our discussions in the end of the penultimate paragraph, section \ref{static}.

\section{Shadows produced by accretion disks\label{ring}}
\subsection{Direct emission, lens ring, and photon ring}
\begin{figure}[htb]
    \centering
    \subfigure[$\ \beta=0.1,\ P=0.6.$]{\includegraphics[width=8cm]{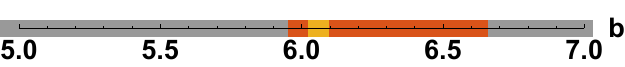}}\\
    \subfigure[$\ \beta=0.2,\ P=0.6.$]{\includegraphics[width=8cm]{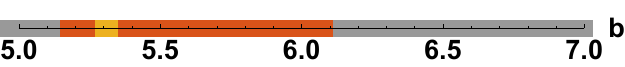}}\\
    \subfigure[$\ \beta=0.2,\ P=0.9.$]{\includegraphics[width=8cm]{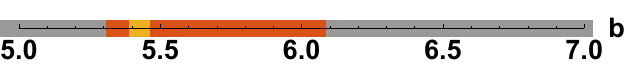}}\\
    \caption{Visualization of Table \ref{NumSol_dir_lens_phtring}. The grey, red, and yellow ranges are direct emission, lens ring, and photon ring, respectively.}
    \label{axes}
\end{figure}

\begin{figure}[htb]
    \centering
    \includegraphics[width=10cm]{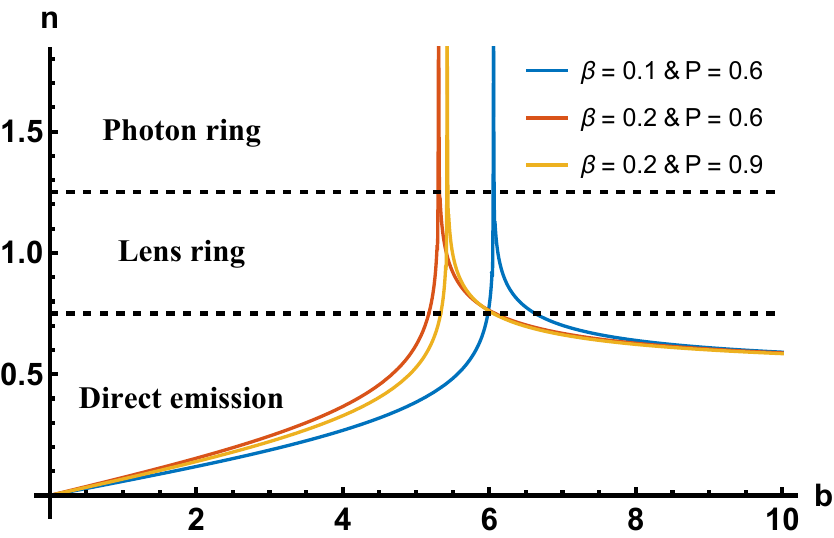}
    \caption{The relationship between the total laps of photons rotating the black hole $n$ and the impact parameter $b$.}
    \label{nb}
\end{figure}

In this section, we study the shadows cast by a thin accretion disk. The accretion disk can be placed on the equatorial plane by defining that the z-axis is perpendicular to the disk plane, for spherically symmetric spacetime. Inspired by \cite{Gralla:2019xty}, the trajectories of photons can be classified into three categories according to the total laps $n$ that orbit the BH. The trajectories with $n<3/4$ are defined as direct emission, where photons intersect the accretion disk only once. The lens ring, where photons pass through the accretion disk twice, corresponds to the orbit's laps within $3/4<n<5/4$. Photons on the photon ring travel across the accretion disk at least three times, satisfying $n>5/4$. The ranges of direct emission, lens ring, and photon ring with respect to the impact parameter $b$ are numerically solved in Table \ref{NumSol_dir_lens_phtring} and Fig. \ref{nb} with different parameters $\beta$ and $P$. Fig. \ref{axes} is a visualization for Table \ref{NumSol_dir_lens_phtring}.
\begin{table}[htbp]
    \setlength{\tabcolsep}{3mm}
    \begin{tabular}{@{}|c|c|c|c|@{}}
    \toprule[1pt]
        & $\beta=0.1$ and $P=0.6$ & $\beta=0.2$ and $P=0.6$ & $\beta=0.2$ and $P=0.9$ \\ \toprule[1pt]
        \begin{tabular}[c]{@{}c@{}}Direct emission\\ $0\leq n<3/4$\end{tabular} & \begin{tabular}[c]{@{}c@{}}$0\leq b<5.98186$\\ $b>6.62840$\end{tabular} & \begin{tabular}[c]{@{}c@{}}$0\leq b<5.17489$\\ $b>6.08053$\end{tabular} & \begin{tabular}[c]{@{}c@{}}$0\leq b<5.33738$\\ $b>6.05728$\end{tabular} \\ \toprule[1pt]
        \begin{tabular}[c]{@{}c@{}}Lens ring\\ $3/4<n<5/4$\end{tabular} & \begin{tabular}[c]{@{}c@{}}$5.98186<b<6.05358$\\ $6.06645<b<6.62840$\end{tabular} & \begin{tabular}[c]{@{}c@{}}$5.17489<b<5.30104$\\ $5.32213<b<6.08053$\end{tabular} & \begin{tabular}[c]{@{}c@{}}$5.33738<b<5.42038$\\ $5.43528<b<6.05728$\end{tabular} \\ \toprule[1pt]
        \begin{tabular}[c]{@{}c@{}}Photon ring\\ $n>5/4$\end{tabular} & $6.05358<b<6.06645$ & $5.30104<b<5.32213$ & $5.42038<b<5.43528$\\ \toprule[1pt]
    \end{tabular}
    \caption{The ranges of direct emission, lens ring, and photon ring with respect to the impact parameter $b$ in the left panels for different parameters $\beta$ and $P$ with fixed $M=1$.}
    \label{NumSol_dir_lens_phtring}
\end{table}

Table \ref{NumSol_dir_lens_phtring}, Figs. \ref{axes} and \ref{nb} present that the range of direct emission extends with the increase of $P$, but the ranges of the lens ring and photon ring both decrease. Interestingly, all these three ranges are shifted outwards with the increase of $P$ or the decrease of $\beta$. Hence, in the presence of magnetic charge, the increase of BI effect obviously strengthens the effect of magnetic charge, which is similar to the result of the blue region of Fig. \ref{risco3rh} in section \ref{isco}. Changes in the BI parameter and the magnetic charge thus have contrary impacts on these three regions.
\begin{figure}[htb]
    \begin{center}
        \subfigure[$\ \beta=0.1,\ P=0.6.$]{\includegraphics[width=5.4cm]{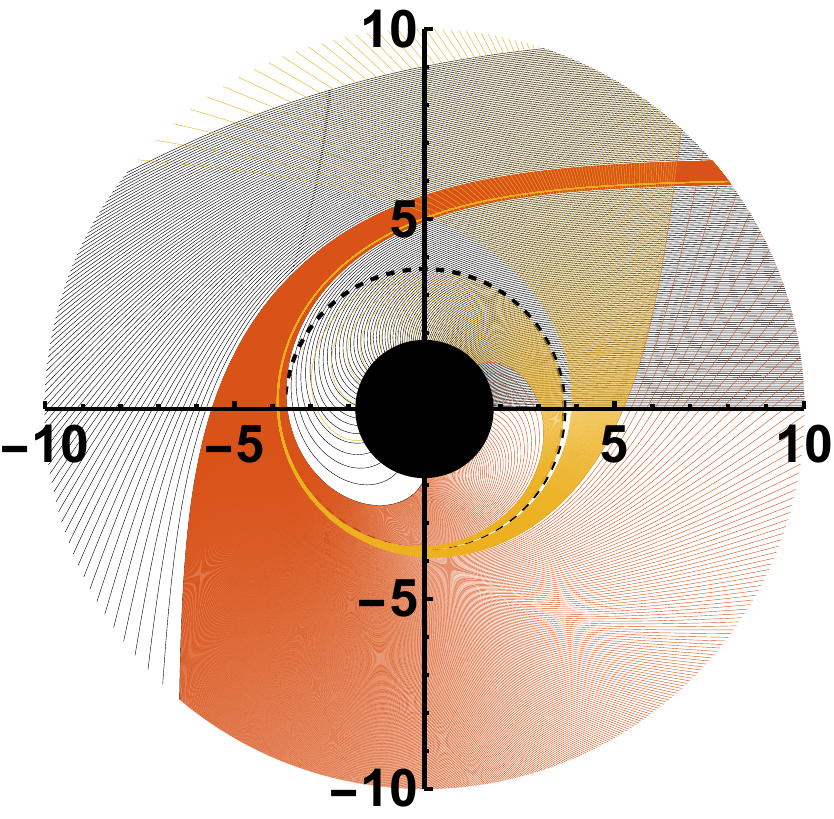}}
        \subfigure[$\ \beta=0.2,\ P=0.6.$]{\includegraphics[width=5.4cm]{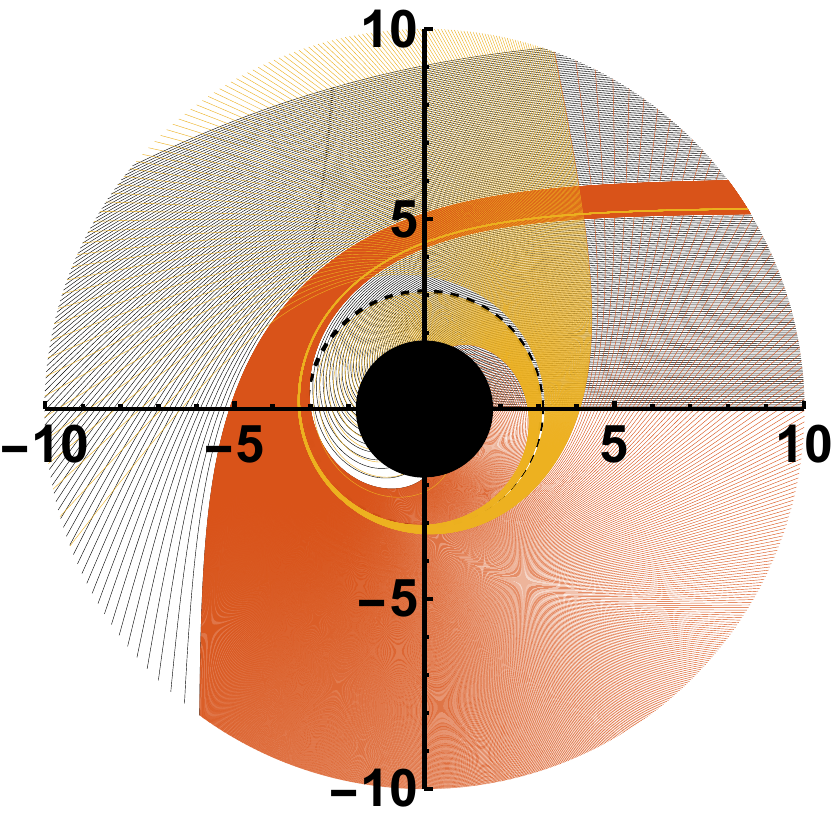}}
        \subfigure[$\ \beta=0.2,\ P=0.9.$]{\includegraphics[width=5.4cm]{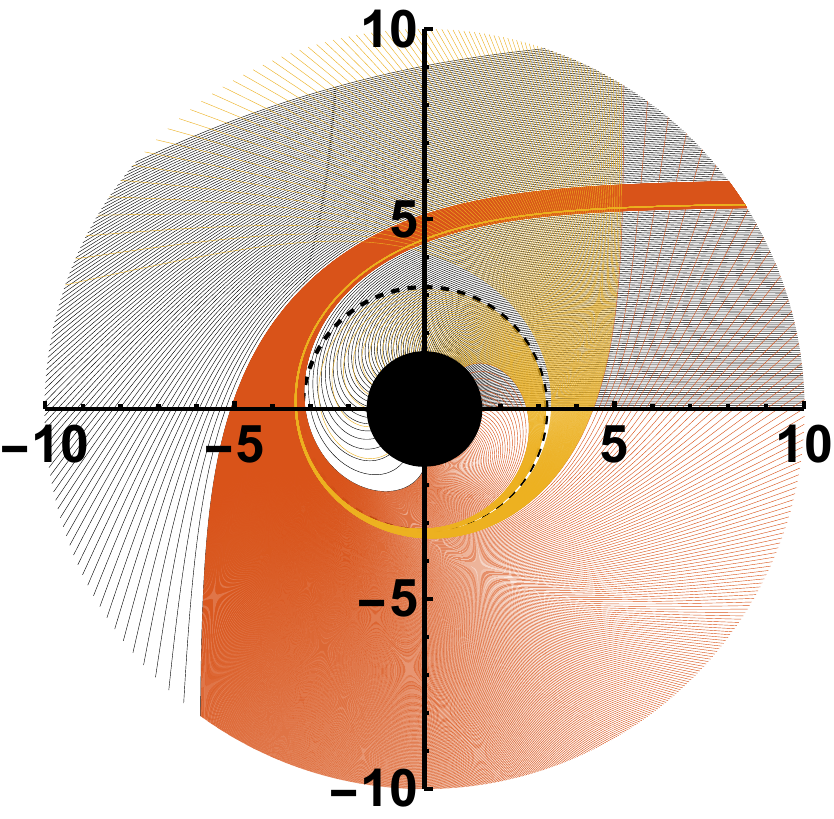}}
    \end{center}
    \caption{Trajectories of photons. The black, red, and yellow curves represent the light rays corresponding to direct emission, lens ring, and photon ring, respectively. And the photon sphere is depicted here by black dashed lines. The black disk represents the BH. The difference in impact parameters between two adjacent trajectories are $0.05$, $10^{-3}$, and $10^{-4}$ in direct emission, lens ring, and photon ring, respectively.}
    \label{geo_DirLenPht}
\end{figure}

We plot the photons' trajectories close to the BH in Fig. \ref{geo_DirLenPht} to illustrate how the magnetic charge and BI effect affect these three ranges. From Fig. \ref{geo_DirLenPht} we note that when $\beta$ is fixed, the increase of $P$ significantly extends the photon ring, and slightly reduces the radius of the event horizon. Meanwhile, the range of the lens ring becomes narrower. If the magnetic charge is fixed, the increase of the BI parameter would expand the range of the lens ring. Interestingly, the impact of weakening BI effect is similar to that of reducing the magnetic charge of BH, which is consistent with our previous analysis.

\subsection{Observed specific intensities and transfer function}
We assume that the thin accretion disk emits isotropically in the rest frame of static worldlines, and the accretion matter is set to be fixed. The accretion disk is placed at the equatorial plane (the y-axis in Fig. \ref{geo_DirLenPht}), and the observer is located at infinity above the north pole ($x=+\infty$ in Fig. \ref{geo_DirLenPht}). The specific intensity received by the observer with emission frequency $\nu_{\text{e}}$ is \cite{Gralla:2019xty}
\begin{align}
    I_{\text{obs}}(r,\nu_{\text{o}})=g^{3}I_{\text{emi}}(r,\nu_{\text{e}}),\qquad g=\sqrt{k(r)f(r)},
    \label{received_intensity_def}
\end{align}
where $I_{\text{emi}}(r,\nu_{\text{e}})$ is the emitted specific intensity by the accretion disk, $\nu_{\text{o}}$ is the observed frequency, and $g$ is the redshift factor derived in Eq. (\ref{g_static_derived}). The total specific intensity $I_{\text{o}}(r)$ can be obtained by integrating all
frequencies of $I_{\text{obs}}(r,\nu_{\text{o}})$, which is denoted as
\begin{align}
    I_{\text{o}}(r)=\int_{0}^{+\infty}I_{\text{obs}}(r,\nu_{\text{o}}){\rm{d}}\nu_{\text{o}}=\int_{0}^{+\infty}g^{4}I_{\text{emi}}(r,\nu_{\text{e}}){\rm{d}}\nu_{\text{e}}=k^{2}(r)f^{2}(r)I_{\text{e}}(r),
    \label{received_intensity_derived}
\end{align}
where we have defined $I_{\text{e}}(r)=\int_{0}^{+\infty}I_{\text{emi}}(r,\nu_{\text{e}}){\rm{d}}\nu_{\text{e}}$ as the total emitting specific intensity.

We assume the disk is optically thin, i.e., no photons are absorbed by the disk. When the photons are emitted from the accretion disk, the light rays are bent by the BH and may intersect with the accretion disk for several times. For the light rays in the regions of direct emission, they intersect with the accretion disk only once. The photons with $3/4<n<5/4$ cross the equatorial plane twice. The light rays intersecting the accretion disk three or more times form the photon ring.

\begin{figure}[htb]
    \begin{center}
        \subfigure[$\ \beta=0.1,\ P=0.6.$]{\includegraphics[width=5.4cm]{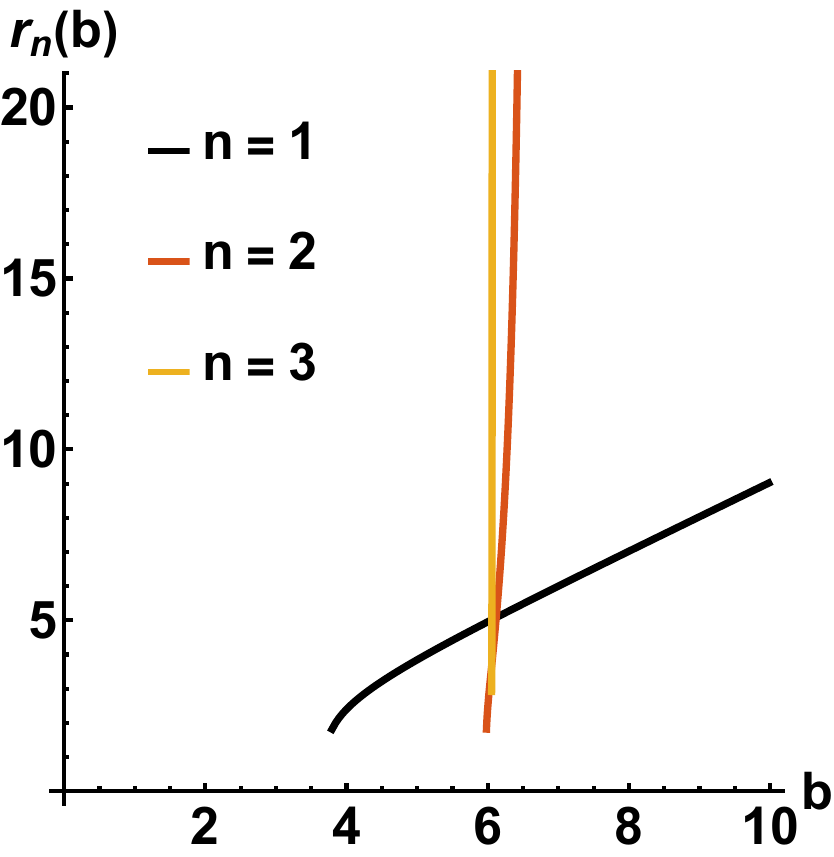}}
        \subfigure[$\ \beta=0.2,\ P=0.6.$]{\includegraphics[width=5.4cm]{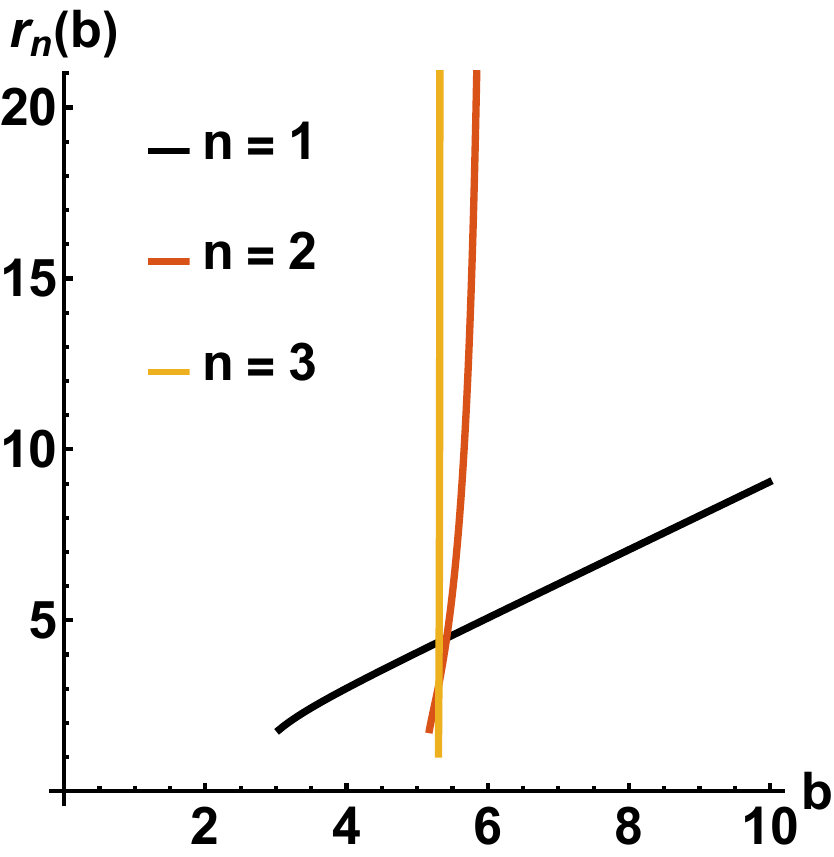}}
        \subfigure[$\ \beta=0.2,\ P=0.9.$]{\includegraphics[width=5.4cm]{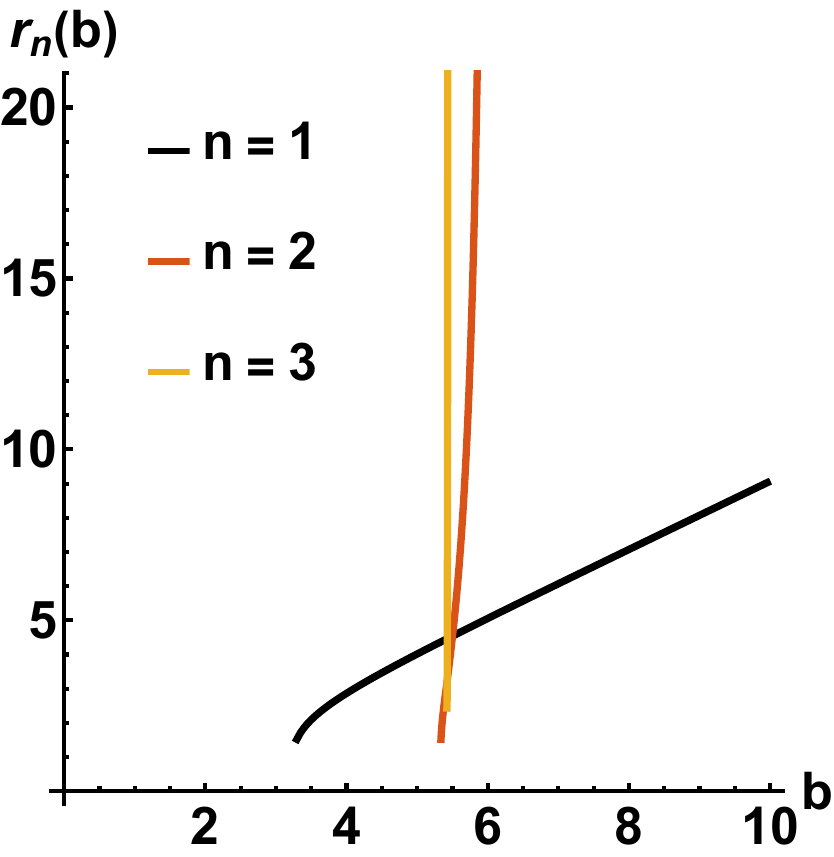}}
    \end{center}
    \caption{The first three transfer functions of a magnetically charged BI BH with a thin accretion disk. The black, red and yellow lines represent the first, second and third transfer functions, respectively.}
    \label{transfunc}
\end{figure}

The sum of the observed specific intensities from each intersection determines the total received intensity, for photons contribute the specific intensity received by the observer once every time they pass through the accretion disk \cite{Li:2021ypw}, yielding
\begin{align}
    I_{\text{o}}(r)=\sum_{n=1}^{\infty}\left.k^{2}(r)f^{2}(r)I_{\text{e}}(r)\right|_{r=r_{n}(b)}.
    \label{received_intensity_sum}
\end{align}
The $r_{n}(b)$ above is the transfer function, where $n$ is a positive integer. It is the radial position where the photons cross the accretion disk for the $n$-th time. The slope of a transfer function is called the demagnification factor \cite{Gralla:2019xty}. At given impact parameter $b$ it represents how much the corresponding part of the accretion disk is demagnified in the observer's vision at infinity. For different $\beta$ and $P$, the first three transfer functions $r_{n}(b)$ are plotted in Figure \ref{transfunc}. Here, we ignore the transfer functions of $n\geq 4$ due to limited numerical calculation precision.

It is noteworthy that the demagnification factors ${\rm{d}}r/{\rm{d}}b$ of these transfer functions are quite different. The first transfer function corresponds to the direct emission, which looks like a straight line with a demagnification factor of 1 approximately. So $r_{1}(b)$ is related to the redshifted source profile. The second transfer function relates to the lens ring (including the photon ring). The slope of $r_{2}(b)$ is much greater than that of $r_{1}(b)$, which indicates the lens ring is a significantly demagnified image of the back side of the accretion disk. The third transfer function is a line almost perpendicular to the horizontal axis and only contains the photon ring. Hence, the image of the photon ring is an extremely demagnified image of the front side of the disk. Thereon, the demagnification factors of further transfer functions approach infinity, and contribute negligibly to the observed intensity. Ignoring the higher-order transfer functions, therefore, has little effect on the total observed intensity.

\subsection{Observational appearances}
\begin{figure}[htb]
    \begin{center}
        \subfigure[\ Exponential model.]{\begin{minipage}{.31\linewidth}
            \includegraphics[width=\linewidth]{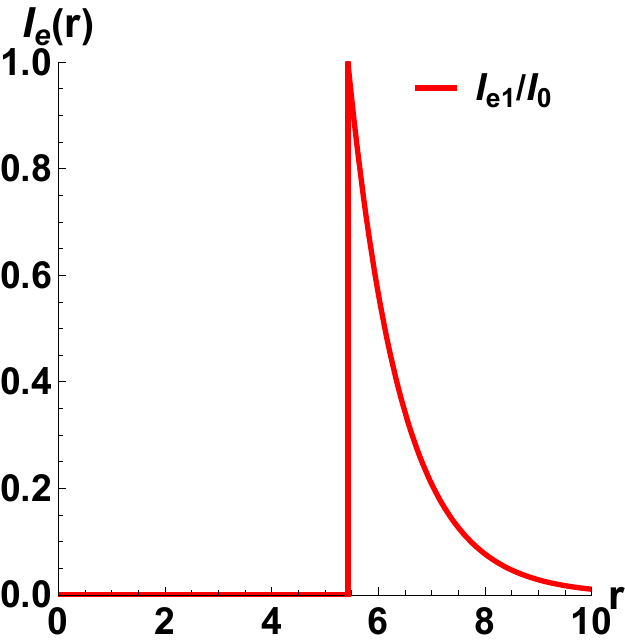}\vspace{2pt}
            \includegraphics[width=\linewidth]{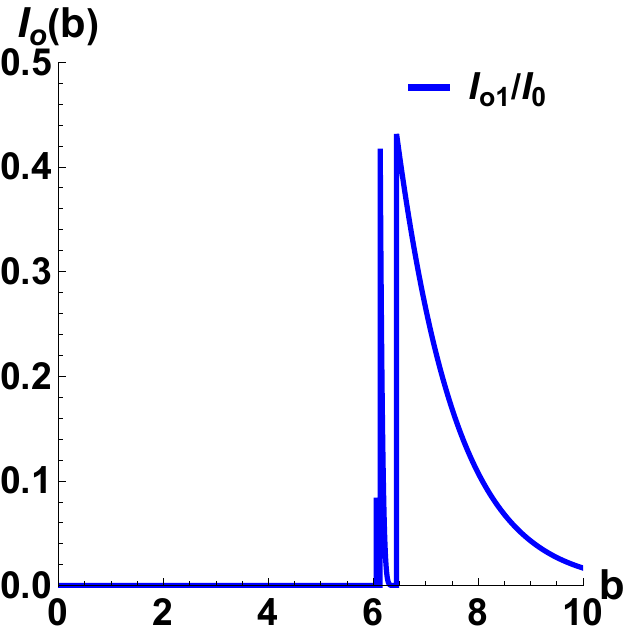}\vspace{2pt}
            \includegraphics[width=\linewidth]{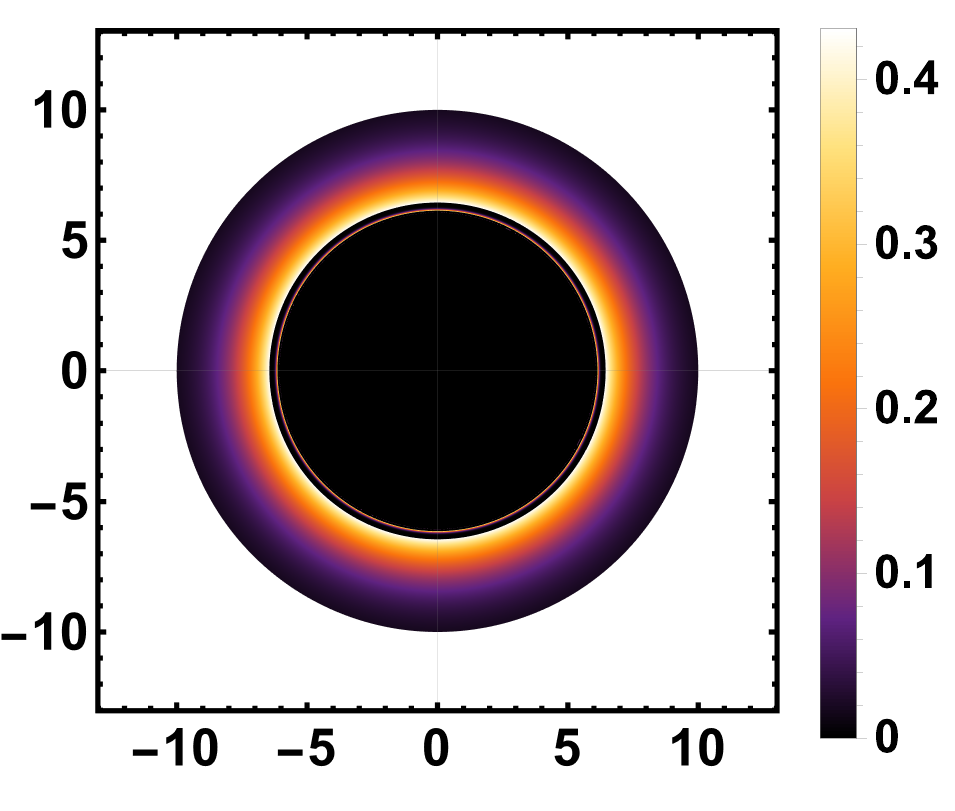}
        \end{minipage}
        }
        \subfigure[\ Power law exponential model.]{\begin{minipage}{.31\linewidth}
            \includegraphics[width=\linewidth]{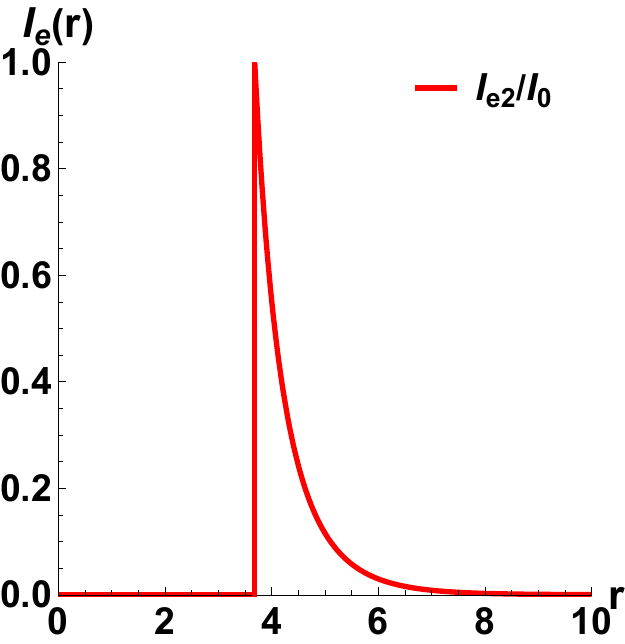}\vspace{2pt}
            \includegraphics[width=\linewidth]{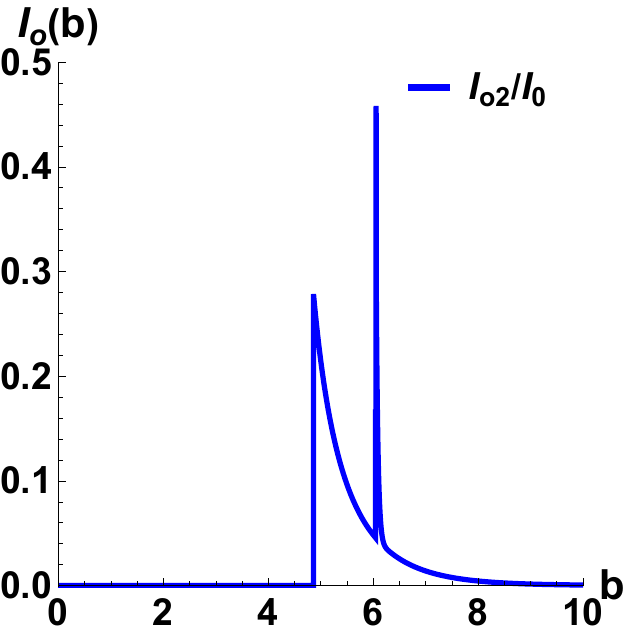}\vspace{2pt}
            \includegraphics[width=\linewidth]{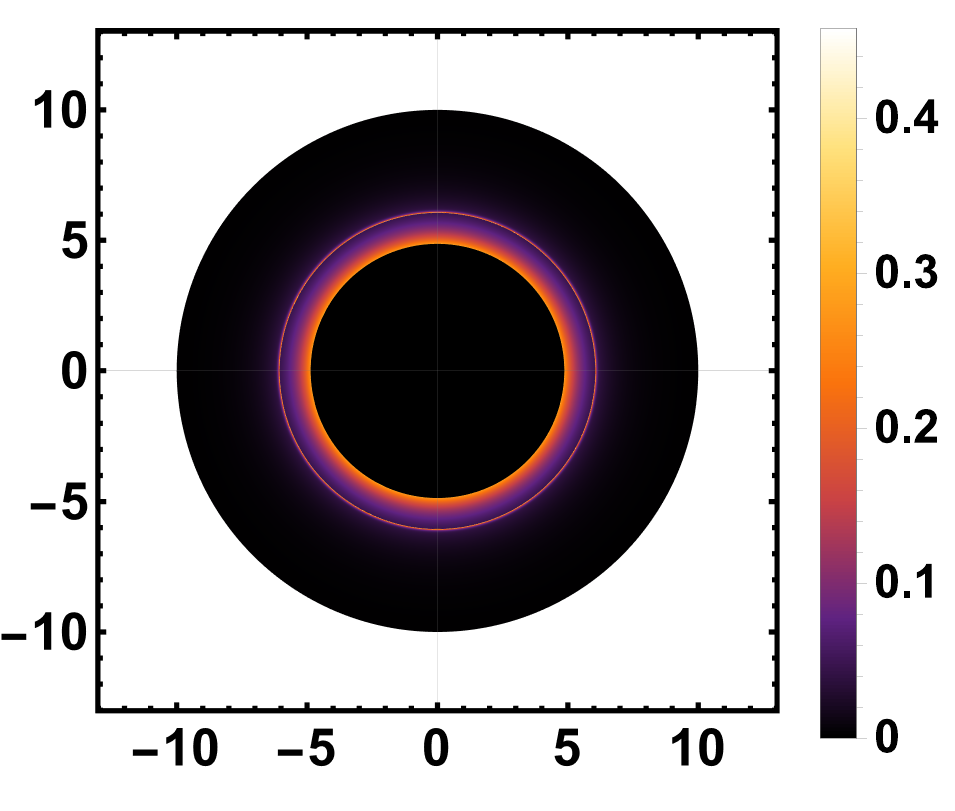}
        \end{minipage}
        }
        \subfigure[\ Bell-shaped model.]{\begin{minipage}{.31\linewidth}
            \includegraphics[width=\linewidth]{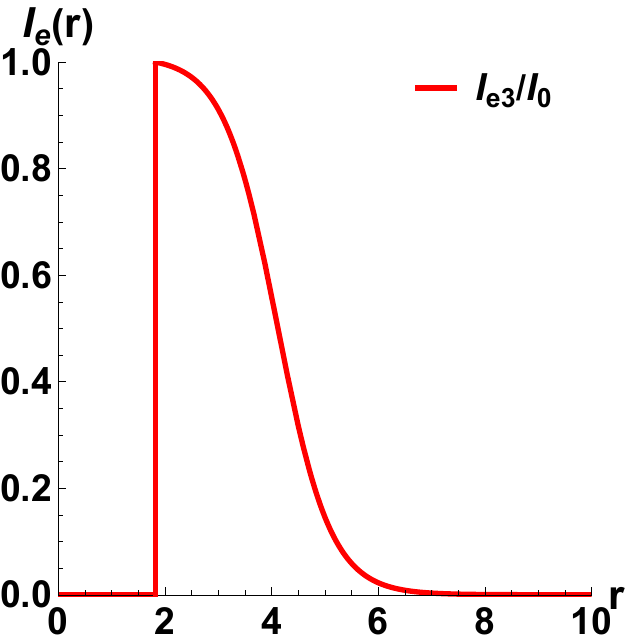}\vspace{2pt}
            \includegraphics[width=\linewidth]{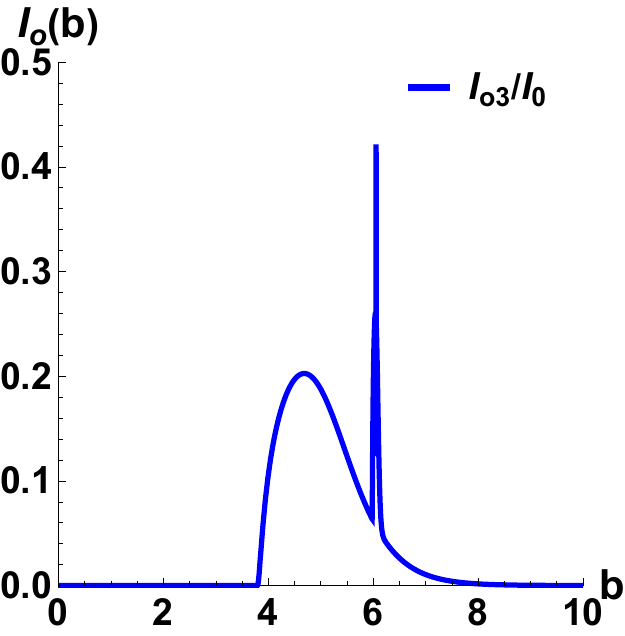}\vspace{2pt}
            \includegraphics[width=\linewidth]{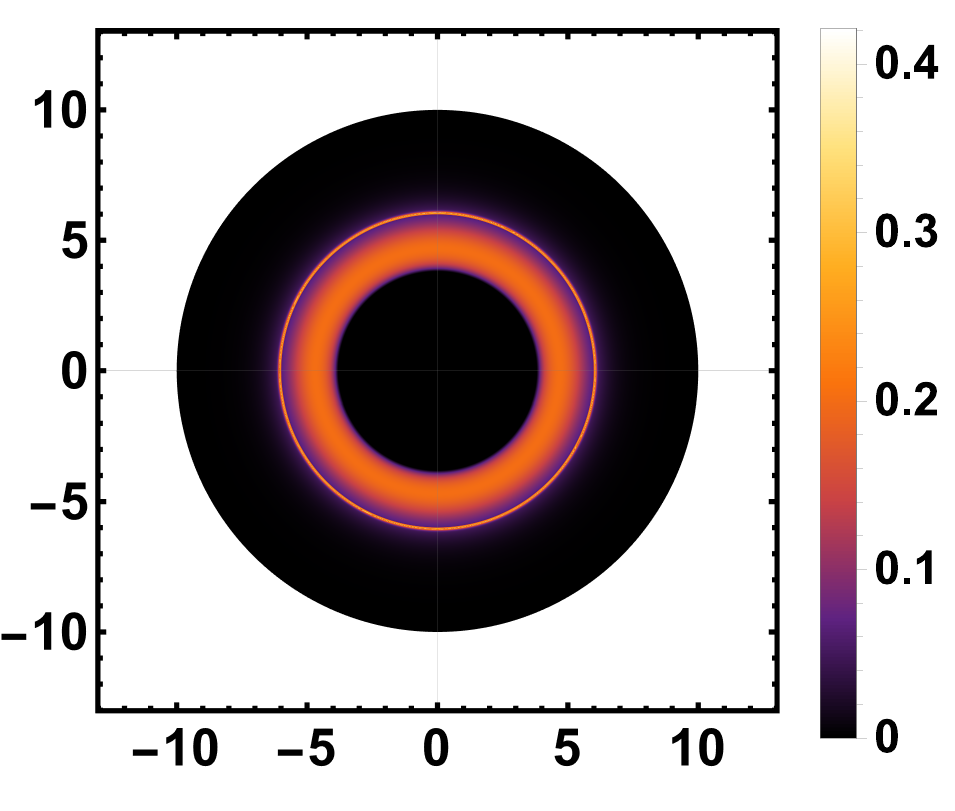}
        \end{minipage}
        }
    \end{center}
    \caption{Observational appearances of a thin disk for different models with fixed $\beta=0.1$ and $P=0.6$. The first row shows the emissive functions $I_{\text{e}}(r)$, and the figures in the second row are the observed intensities $I_{\text{o}}(b)$. In the third row, we plot these observed intensities into two-dimensional disks.}
    \label{shadow_ring_beta0.1_P0.6}
\end{figure}

\begin{figure}[htb]
    \begin{center}
        \subfigure[\ Exponential model.]{\begin{minipage}{.31\linewidth}
            \includegraphics[width=\linewidth]{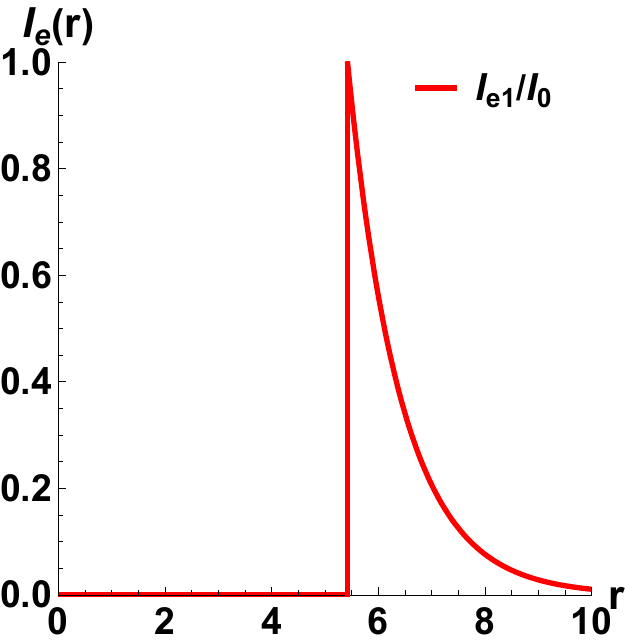}\vspace{2pt}
            \includegraphics[width=\linewidth]{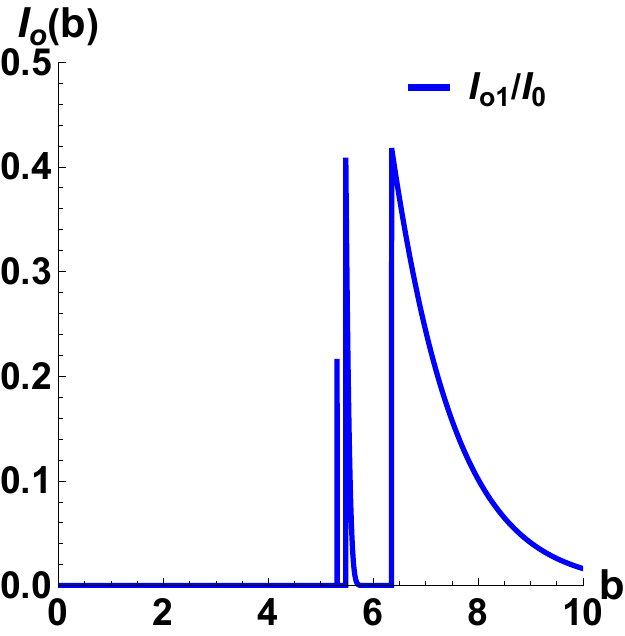}\vspace{2pt}
            \includegraphics[width=\linewidth]{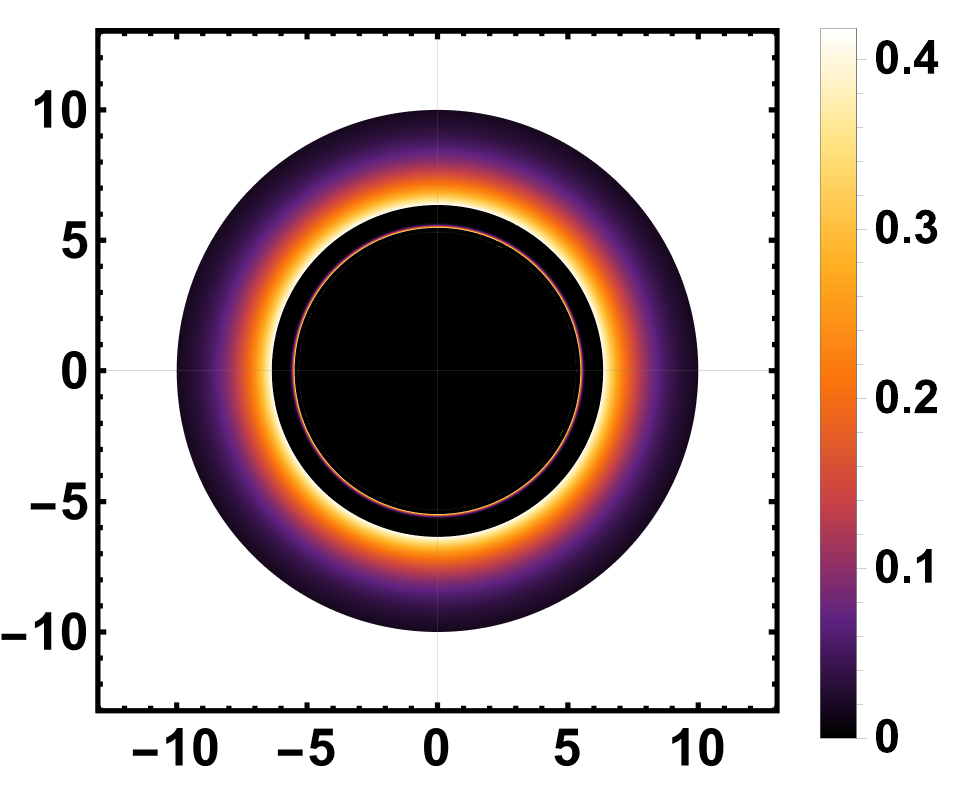}
        \end{minipage}
        }
        \subfigure[\ Power law exponential model.]{\begin{minipage}{.31\linewidth}
            \includegraphics[width=\linewidth]{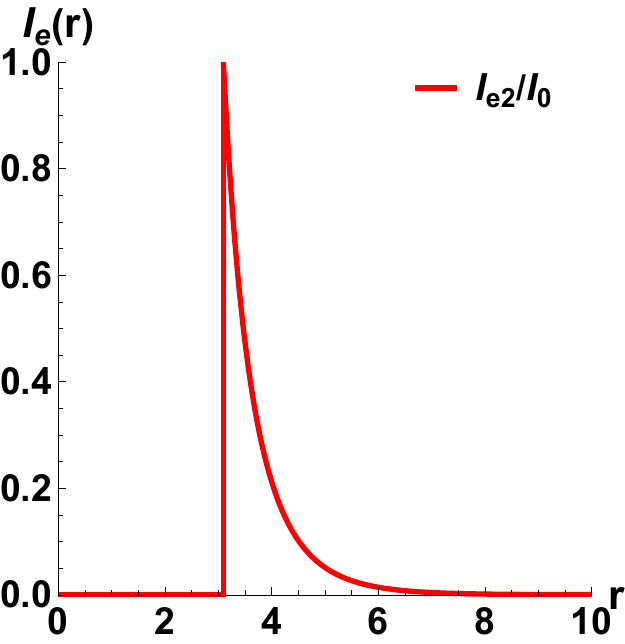}\vspace{2pt}
            \includegraphics[width=\linewidth]{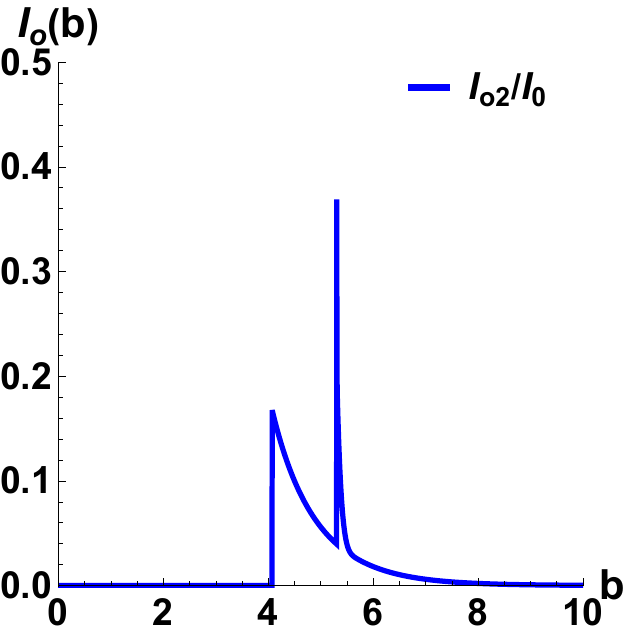}\vspace{2pt}
            \includegraphics[width=\linewidth]{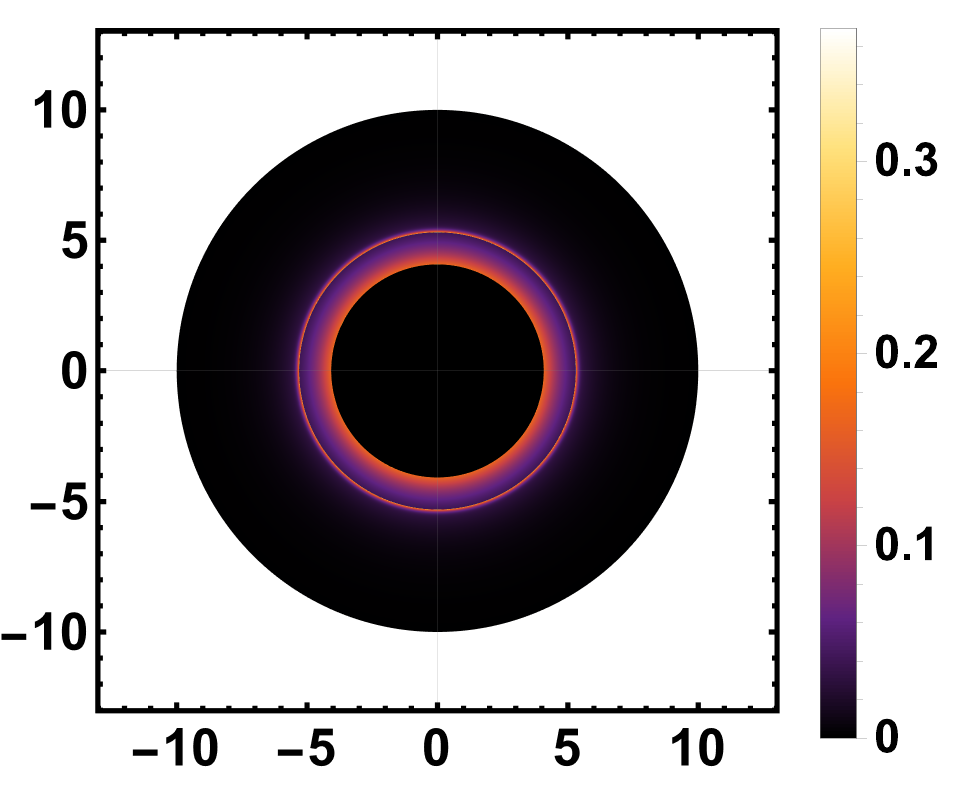}
        \end{minipage}
        }
        \subfigure[\ Bell-shaped model.]{\begin{minipage}{.31\linewidth}
            \includegraphics[width=\linewidth]{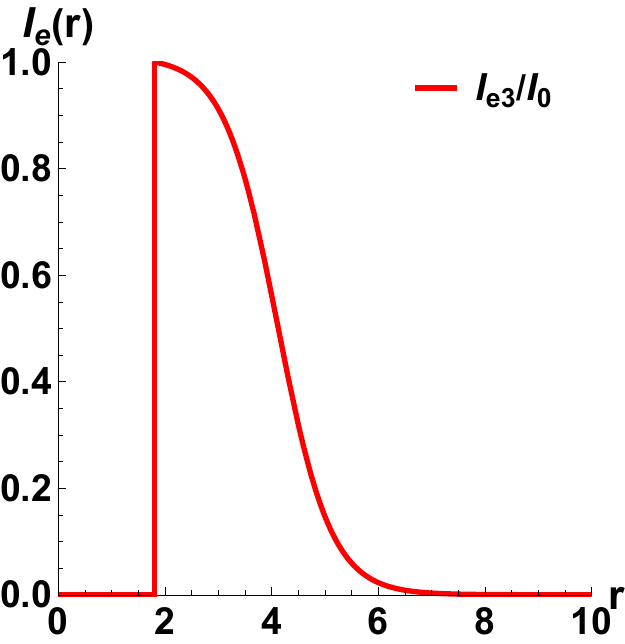}\vspace{2pt}
            \includegraphics[width=\linewidth]{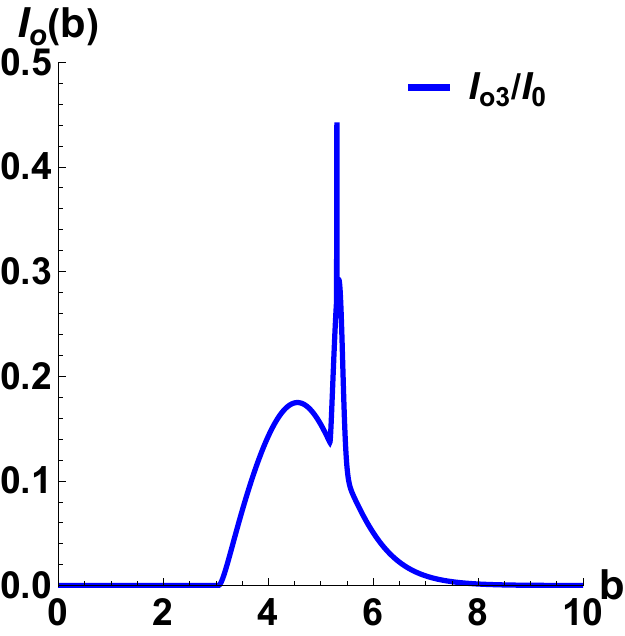}\vspace{2pt}
            \includegraphics[width=\linewidth]{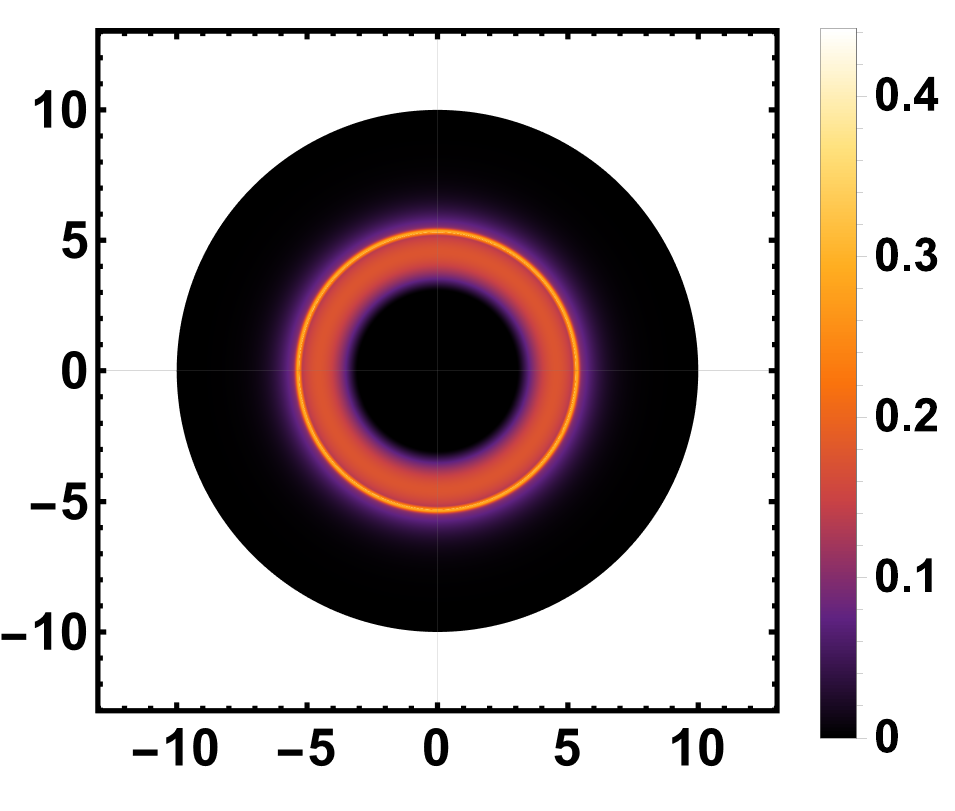}
        \end{minipage}
        }
    \end{center}
    \caption{Observational appearances of a thin disk for different models with fixed $\beta=0.2$ and $P=0.6$. The first row shows the emissive functions $I_{\text{e}}(r)$, and the figures in the second row are the observed intensities $I_{\text{o}}(b)$. In the third row, we plot these observed intensities into two-dimensional disks.}
    \label{shadow_ring_beta0.2_P0.6}
\end{figure}

\begin{figure}[htb]
    \begin{center}
        \subfigure[\ Exponential model.]{\begin{minipage}{.31\linewidth}
            \includegraphics[width=\linewidth]{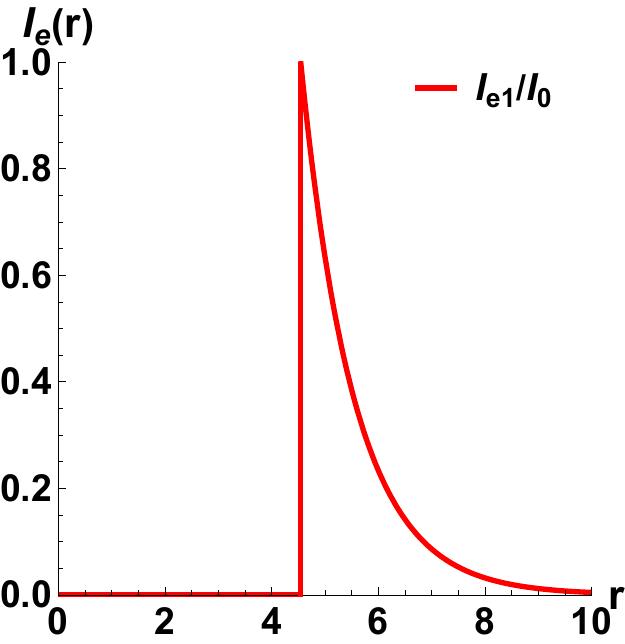}\vspace{2pt}
            \includegraphics[width=\linewidth]{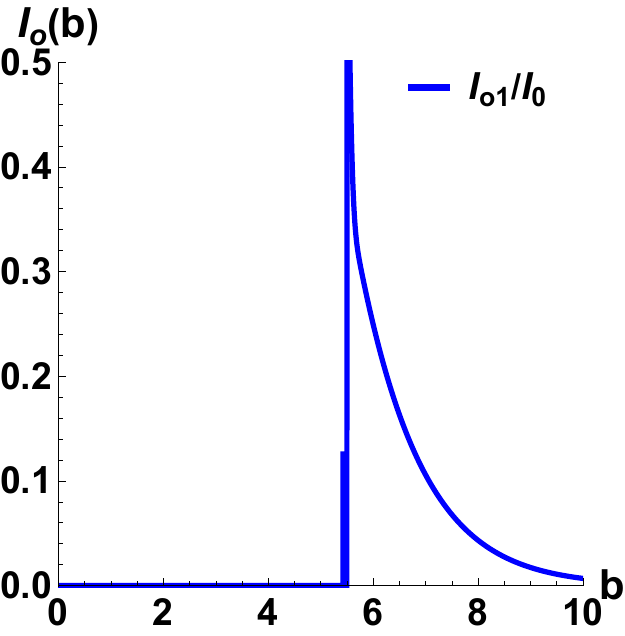}\vspace{2pt}
            \includegraphics[width=\linewidth]{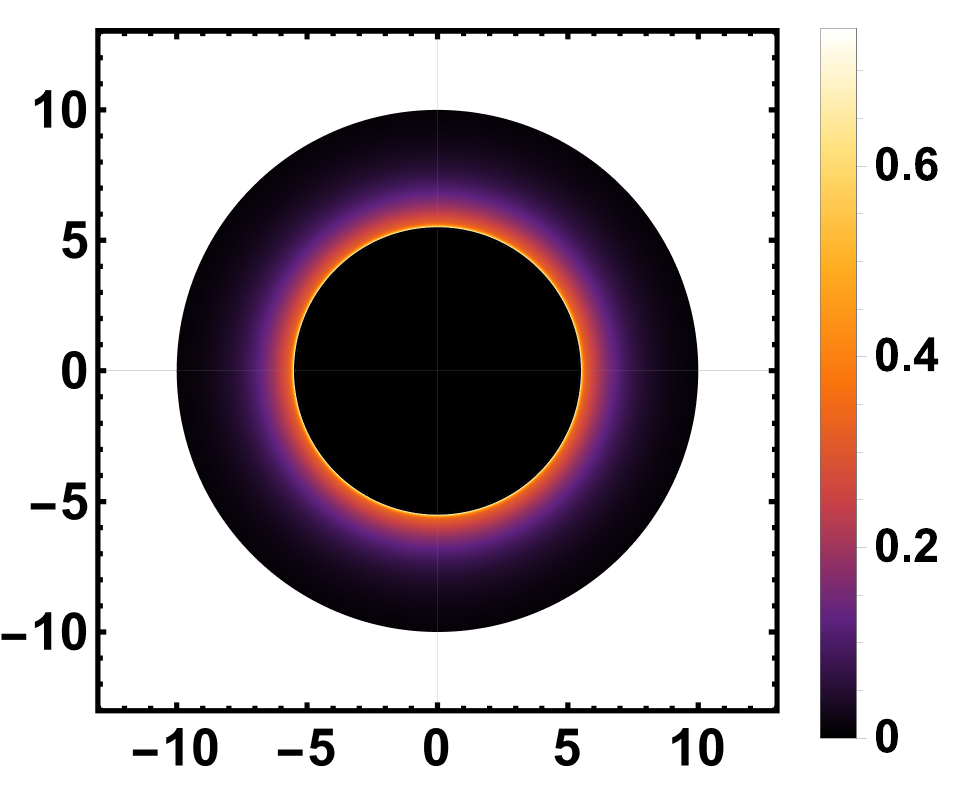}
        \end{minipage}
        }
        \subfigure[\ Power law exponential model.]{\begin{minipage}{.31\linewidth}
            \includegraphics[width=\linewidth]{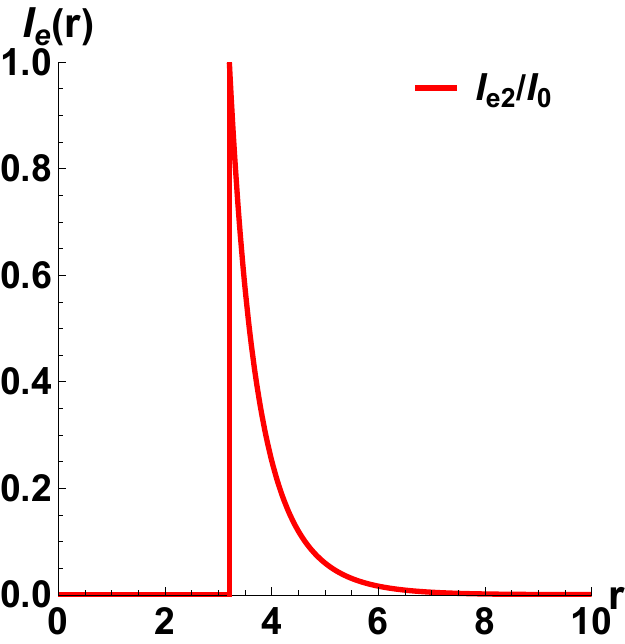}\vspace{2pt}
            \includegraphics[width=\linewidth]{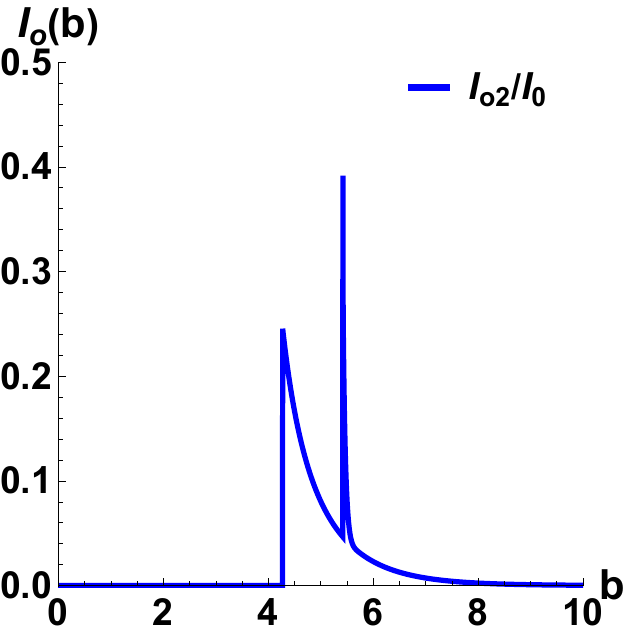}\vspace{2pt}
            \includegraphics[width=\linewidth]{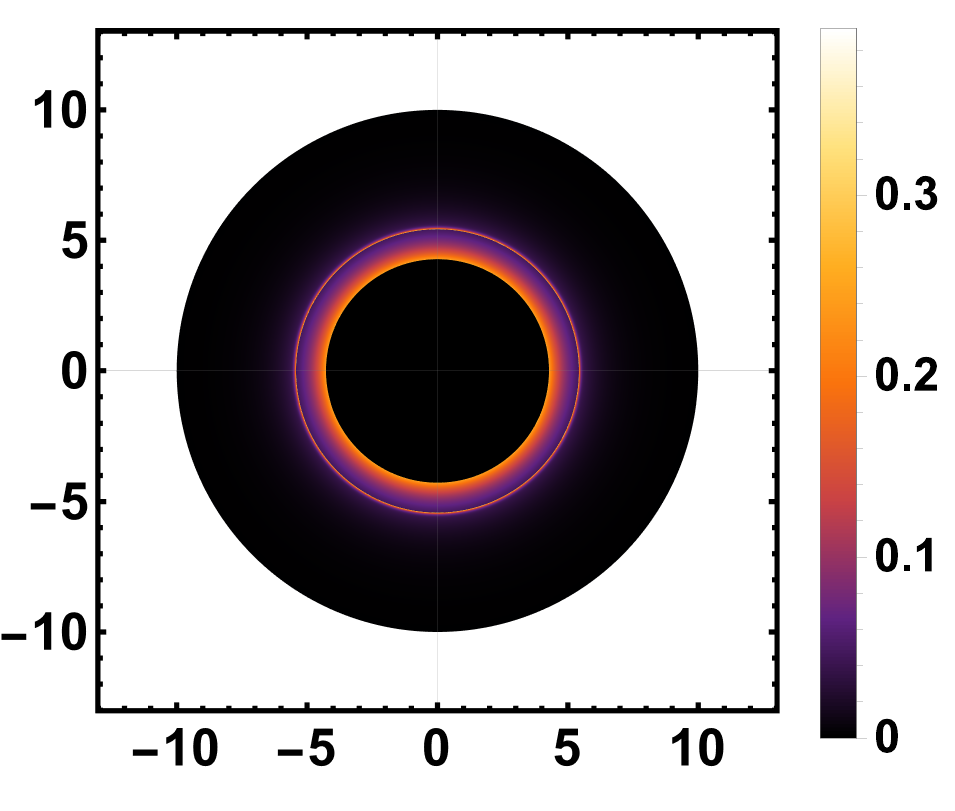}
        \end{minipage}
        }
        \subfigure[\ Bell-shaped model.]{\begin{minipage}{.31\linewidth}
            \includegraphics[width=\linewidth]{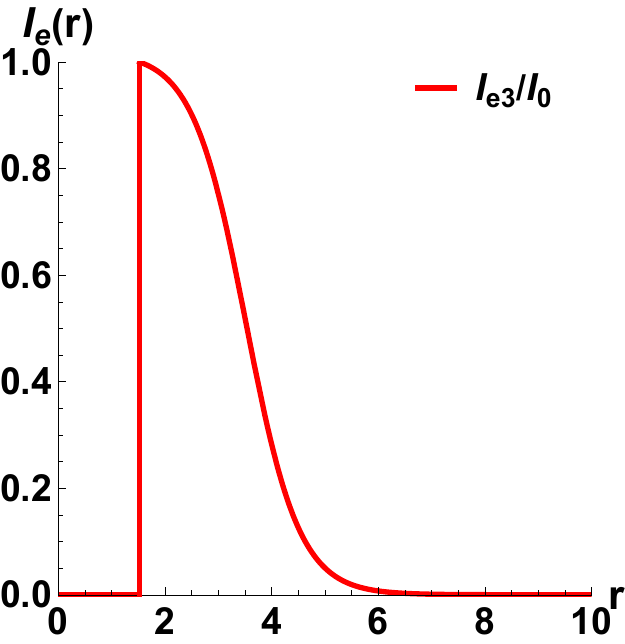}\vspace{2pt}
            \includegraphics[width=\linewidth]{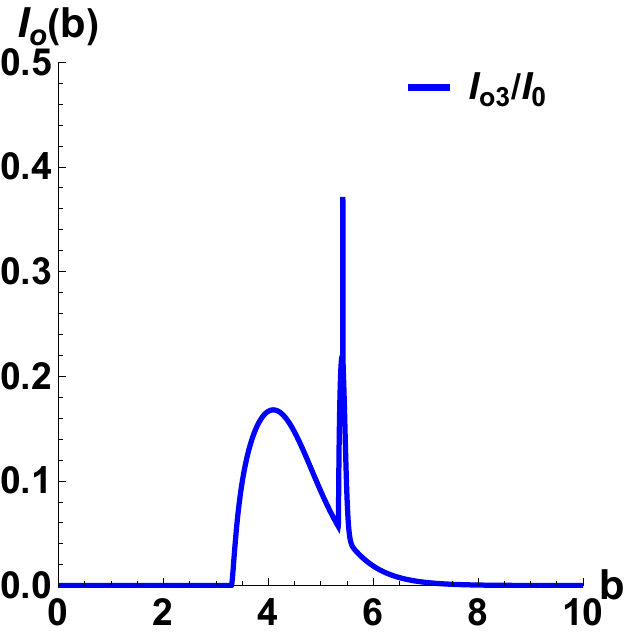}\vspace{2pt}
            \includegraphics[width=\linewidth]{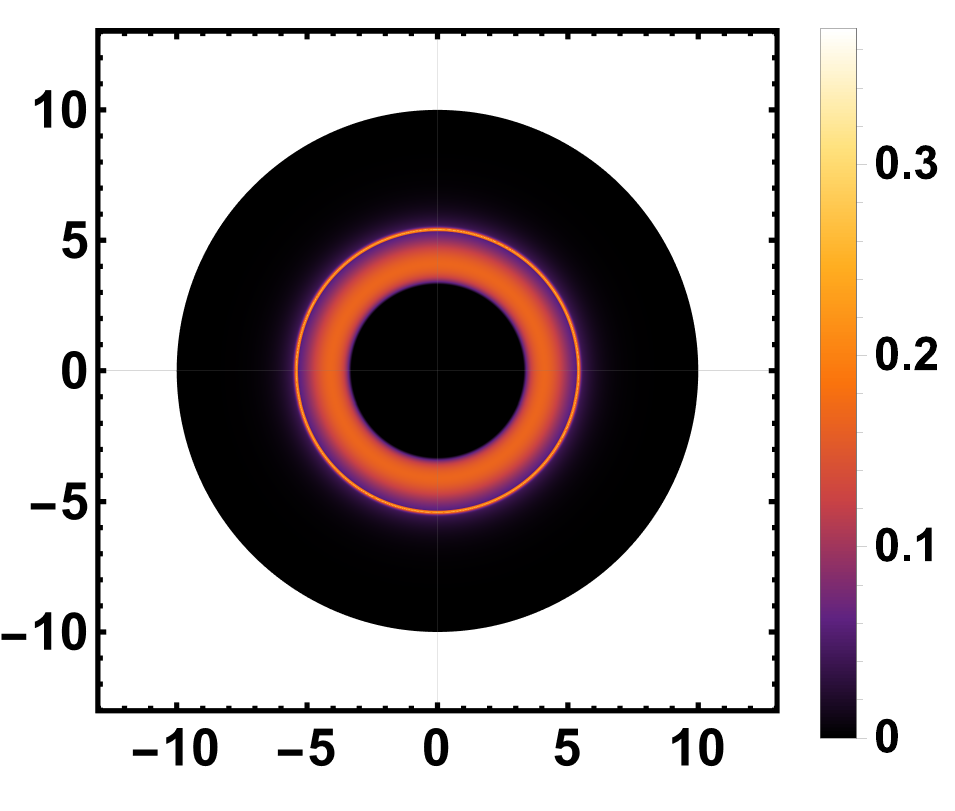}
        \end{minipage}
        }
    \end{center}
    \caption{Observational appearances of a thin disk for different models with fixed $\beta=0.2$ and $P=0.9$. The first row shows the emissive functions $I_{\text{e}}(r)$, and the figures in the second row are the observed intensities $I_{\text{o}}(b)$. In the third row, we plot these observed intensities into two-dimensional disks.}
    \label{shadow_ring_beta0.2_P0.9}
\end{figure}

In this subsection, we study the specific intensity with some numerical toy models based on Eq. (\ref{received_intensity_sum}). The first emissive function $I_{\text{e1}}$ is assumed to be an exponential distribution that starts from $r=r_{\text{ISCO}}$.
\begin{align}
    I_{\text{e1}}(r)/I_{0}=\begin{cases}
        \exp\left[-(r-r_{\text{ISCO}})\right],\ r\geq r_{\text{ISCO}};\\
        0,\ r<r_{\text{ISCO}};
    \end{cases}
    \label{emitted_intensity1}
\end{align}
where $I_{0}$ is the maximum intensity, similarly hereinafter. It is plotted in the top-left figures in Figs. \ref{shadow_ring_beta0.1_P0.6}, \ref{shadow_ring_beta0.2_P0.6}, and \ref{shadow_ring_beta0.2_P0.9} for different parameters $\beta$ and $P$. It reaches its peak at $r=r_{\text{ISCO}}$, and plummets to zero as $r<r_{\text{ISCO}}$. 

The second emissive function $I_{\text{e2}}$ is an inversely proportional function multiplied by an exponential distribution, which starts from $r=r_{\text{ph}}$. We call it power law exponential model for convenience in the following text.
\begin{align}
    I_{\text{e2}}(r)/I_{0}=\begin{cases}
        \frac{1}{r-r_{\text{ph}}+1}\exp\left[-(r-r_{\text{ph}})\right],\ r\geq r_{\text{ph}};\\
        0,\ r<r_{\text{ph}}.
    \end{cases}
    \label{emitted_intensity2}
\end{align}
This emissive function is plotted in the top-middle figures in Figs. \ref{shadow_ring_beta0.1_P0.6}, \ref{shadow_ring_beta0.2_P0.6}, and \ref{shadow_ring_beta0.2_P0.9} for different parameters $\beta$ and $P$. Obviously, it has a peak at $r=r_{\text{ph}}$ and drops steeper than the first emissive function as $r>r_{\text{ph}}$.

The third emissive function is assumed to be a bell-shaped function starting from $r=r_{\text{h}}$. 
\begin{align}
    I_{\text{e3}}(r)/I_{0}=\begin{cases}
        \frac{1-\tanh \left[r-(r_{\text{ISCO}}-r_{\text{h}}+0.5)\right]}{1-\tanh \left[r_{\text{h}}-(r_{\text{ISCO}}-r_{\text{h}}+0.5)\right]},\ r\geq r_{\text{h}};\\
        0,\ r<r_{\text{h}}.
    \end{cases}
    \label{emitted_intensity3}
\end{align}
The third emissive function is plotted in the top-right figures in Figs. \ref{shadow_ring_beta0.1_P0.6}, \ref{shadow_ring_beta0.2_P0.6}, and \ref{shadow_ring_beta0.2_P0.9} for different parameters $\beta$ and $P$. After reaching its maximum at $r=r_{\text{h}}$, it decreases slowly at first and then rapidly. By employing Eq. (\ref{received_intensity_sum}) to these emissive functions, we numerically calculate the observed intensities, which are then plotted in the middle row in those figures corresponding to these three emissive functions.

The panels in the second row of Figs. \ref{shadow_ring_beta0.1_P0.6}, \ref{shadow_ring_beta0.2_P0.6}, and \ref{shadow_ring_beta0.2_P0.9} correspond to the observed intensities of different parameters and emissive functions. For the exponential model, the corresponding impact parameter for the first peak of the observed intensity is about $b\approx 6.1$ for $\beta=0.1$ and $P=0.6$ ($b\approx 5.3$ for $\beta=0.2$ and $P=0.6$; $b\approx 5.5$ for $\beta=0.2$ and $P=0.9$), which is located at the photon ring. The lens ring causes the second, considerably higher peak that occurs immediately after the first one. In the middle-left panels of Figs. \ref{shadow_ring_beta0.1_P0.6} and \ref{shadow_ring_beta0.2_P0.6}, the observed intensities reach their maximum after the second peak. For $\beta=0.1$ and $P=0.6$, the third peak of the observed intensity's corresponding impact parameter is approximately $b\approx 6.6$ ($b\approx 6.1$ for $\beta=0.2$ and $P=0.6$), representing the peak observed intensity produced by direct emission. In the case of $\beta=0.2$ and $P=0.9$, there are only two peaks in the observed intensity, but the second peak is higher than those in the other cases. The reason is that the lens ring is merged with the peak corresponding to the direct emission. In the bottom-left panels of Figs. \ref{shadow_ring_beta0.1_P0.6} and \ref{shadow_ring_beta0.2_P0.6} we can find a narrow ring just within the brightest ring, which is the lens ring. One may zoom in to find the extremely thin photon ring just within the lens ring. The brightness and width of the whole shadow are dominated by direct emission in the exponential model.

For the power law exponential model, all the observed intensities shoot up to their first peak at $b\approx 4.9$, $b\approx 4.1$, and $b\approx 4.3$ in the central panel of Figs. \ref{shadow_ring_beta0.1_P0.6}, \ref{shadow_ring_beta0.2_P0.6}, and \ref{shadow_ring_beta0.2_P0.9} respectively. These local maximums are related to direct emission. The observed intensities then steadily decline until they reach the maximum produced by the combination of the photon ring and the lens ring. The bottom-center panels of those three figures show that the lens ring and photon ring are combined to generate a halo with a highly luminous but extremely narrow profile. And the direct emission has a low observed intensity but a wide range. Among these figures, the lens ring and the photon ring are indistinguishable. For this model, the brightness of the whole shadow is dominated by the combination of the photon ring and the lens ring, but the direct emission contributes the most to the width.

In the middle-right panels of Figs. \ref{shadow_ring_beta0.1_P0.6}, \ref{shadow_ring_beta0.2_P0.6}, and \ref{shadow_ring_beta0.2_P0.9}, three peaks appear from left to right. The luminosity climbs to the first maximum produced by the direct emission. After the first peak, the observed intensities drop gradually except for a dramatic rise to their maximum and fall immediately. The maximum is approximately $b\approx 6.1$ for $\beta=0.1$ and $P=0.6$ ($b\approx 5.3$ for $\beta=0.2$ and $P=0.6$; $b\approx 5.5$ for $\beta=0.2$ and $P=0.9$), which corresponds to the lens ring. The photon ring is a lower peak very close to the lens ring. Although we can identify the lens ring and the photon ring through $I_{\text{o3}}(b)$ curves, it is hard to distinguish them in the simulated shadow images. As a result, the lens ring and photon ring produce a thin and bright halo, and the direct emission contributes a relatively dark but wide ring. For the bell-shaped model, the lens ring, combined with the photon ring, dominates the maximum brightness of the whole shadow, while the direct emission contributes most to the width.

Combining Figs. \ref{shadow_ring_beta0.1_P0.6}, \ref{shadow_ring_beta0.2_P0.6}, and \ref{shadow_ring_beta0.2_P0.9}, we arrive at some interesting conclusions. Firstly, it is easy to notice the differences in the shadows of magnetically charged BI BHs cast by the accretion disks starting from various radii. The central dark area's radius strongly correlates with the model. Secondly, the BI effect and the magnetic charge both have a considerable effect on the shadows. It is noticeable that the increase of the magnetic charge intensifies the luminosity produced by direct emission apparently, and the increase of the BI parameter weakens the effect of the magnetic charge. Lastly, the third is the most consistent with astronomical observations.

Different emission models have a large impact on the observed intensity of direct emission but have little impact on the lens ring and the photon ring. This is due to the small range of the impact parameter related to the lens ring and photon ring. The range of the radial distance from the BH of photons emitted within the aforementioned range of the impact parameter is also narrow. The shapes and locations of the lens ring and the photon ring are almost unchanged with the emission models, for the emissivity of the accretion disk is almost constant in such a short range of the radial distance. The direct emission, however, relates to a wide range of the impact parameter. Photons with an impact parameter in this range have a wide range of radial distance from the BH when emitted from the accretion disk. The shapes and locations of observed intensity corresponding to the direct emission are also changing because the emission intensity of the accretion disk varies with the change in radial distance. Actually, the lens ring and the photon ring of the three models are difficult to identify, for the ranges of the impact parameter corresponding to the transfer functions of $n=2$ and $n=3$ are very narrow. In this case, regardless of the accretion model, the emission intensity is nearly constant in such a narrow range of radial distance. Therefore, it is difficult to distinguish the photon ring and the lens ring only by changing the emission model.

\section{Conclusion and discussion\label{concl}}
In this paper, we investigated the observational appearances of magnetically charged Born-Infeld black hole. Firstly we presented the metric for BI BH with magnetic charge, and studied the existence of black holes with unit mass in the parameter space of the BI parameter $\beta$ and magnetic charge $P$.  The critical value of the magnetic charge increases with the increasing effect of BI nonlinear electrodynamics. The radii of the event horizon, the innermost stationary circular orbit and the photon sphere, i.e., $r_{\text{h}}$, $r_{\text{ISCO}}$, and $r_{\text{ph}}$, and their corresponding impact parameter $b_{\text{ph}}$ decrease as $\beta$ increases. $r_{\text{ph}}$ and $r_{\text{ISCO}}$ monotonically decrease as $P$ increases, but $r_{\text{ph}}$ and $b_{\text{ph}}$ increase consistently. 

To compare the spacetime of a magnetically charged BI BH with that of the Schwarzschild BH, we further investigated the values of $r_{\text{ISCO}}-3r_{\text{h}}$ and found that $\beta$ has critical values for each magnetic charge. The value $r_{\text{ISCO}}-3r_{\text{h}}$ increases monotonically as the BI effect weakens, i.e., the increase of $\beta$. But the increase of the magnetic charge decreases the aforementioned value if $\beta$ is below the critical value, and vice versa. The geodesics are then derived from the equivalent metric of photons. The photon trajectories are null geodesics in the effective metric rather than the background metric because of the effects of BI. We found that the photon geodesics bent more as the BI parameter decreases or the magnetic charge increases. 

BH shadows are produced by photons that are deflected in curved spacetime. We first consider static spherically symmetric accretion, assuming that the density of photons follows a lognormal distribution with a maximum at the radius of the photon sphere. Also, we supposed that the frequency of the photons emitted is not monochromatic but has a normal distribution with central frequency $\nu_{\text{c}}$. The specific intensity received by a distant observer was obtained by integrating along the geodesics using the backward ray tracing method. And we plotted the specific intensity on a two-dimensional plane with the observer located at infinity above the north pole to display the BH shadows. The intensity grows slightly as $b<b_{\text{ph}}$, then rises sharply to the maximum as $b$ approaches $b_{\text{ph}}$, and asymptotically drops to zero when $b>b_{\text{ph}}$. The specific intensity reaches the maximum when the impact parameter of photons is equal to the critical value, as a result of the photons orbiting the BH infinite times. 

We also took infalling spherically symmetric accretion models into account. The shadows of infalling accretion are obviously darker than the static ones with the same BI parameter $\beta$ and magnetic charge $P$ because the impact of the redshift factor $g$ on specific intensity becomes significant when the photons get close to the BH. The brightest ring in each shadow represents its photon sphere in both infalling and static accretion. The radii and intensities of the shadows vary depending on the BI parameter and the magnetic charge. It is also noteworthy that the impact parameter $b$ corresponding to the peak of the specific intensity is always the same in both static and infalling accretions, indicating that the critical value $b_{\text{ph}}$ is only related to the effective metric. Moreover, the radius of the photon sphere is independent of the accretions. 

We then studied the shadow produced by accretion disks. The photons emitted by the disk are divided into direct emission, lens ring, and photon ring based on their numbers of intersections with the disk. The regions of direct emission, lens ring, and photon ring with parameter changes are similar to the previous analysis of $r_{\text{ISCO}}$. We introduced transfer functions to describe the relationship between the radial positions where the photons intersect the accretion disk and their impact parameters. From the demagnification factors of the transfer functions, we illustrated that the slope of direct emission is almost fixed. However, the lens ring and the photon ring are noticeably demagnified. Then the observational appearances are simulated using three simple emission models of the accretion disk. Among all the simulations, the direct emission dominates the width of the shadow. The photon ring and the lens ring can hardly be distinguished only by changing the emission model.

Moreover, we compare our findings to regular BH in rational NLED in Appendix \ref{appendixComparison}. The comparison of regular magnetic BH in rational NLED with magnetically charged BI BH is summarized in Table \ref{comparison}. Observational appearances of regular BH in rational NLED with different magnetic charge and rational NLED parameter $\beta_r$ were also investigated. We found that the observational appearances of a magnetic BH with spherical accretions are inferred to be similar to the one for BI BH.

With increasing resolutions of astrophysical observations, the nuances between the inner radii of shadows and the theoretical solutions for $r_{\text{ISCO}}$ may be found, implying the existence of magnetic charge combined with the BI NLED is probable. We expect this paper to provide a reference for the investigation of shadows of BI BH and the effect of magnetic fields on BI BH.

\begin{acknowledgments}
    The authors are grateful to Peng Wang, Aoyun He, and Yadong Xue for useful discussions. And the authors would like to thank the anonymous referee for helpful comments and suggestions, which helped to improve the quality of this paper. This work is supported by NSFC Grants No. 12147207, 12175212, and 12275184.
\end{acknowledgments}

\appendix
\begin{appendices}
\section{Comparison between regular BH in rational NLED and magnetically charged BI BH\label{appendixComparison}}
It was shown in Ref. \cite{Bronnikov:2000vy} that magnetically charged BH can have regular solution when NLED becomes Maxwell electrodynamics at weak fields. And a good example is the rational NLED models which has regular BH solutions without singularity at center \cite{Kruglov:2020tes,Kruglov:rationalNLED}. In this Appendix we compare our findings to regular BH in rational NLED. Following \cite{Kruglov:2017mpj}, we introduce magnetic energy density by combining Eqs. (\ref{LF}), (\ref{EngMomTensor}), and (\ref{EMfield})
\begin{align}
    \rho_{M}=T_{0}^{0}=L(F)=4\beta^{2}\left( 1-\sqrt{1+\frac{F}{2\beta^{2}}} \right).
\end{align}
We attempt to introduce the magnetic mass
\begin{align}
    m_{M}=\int_{0}^{+\infty}\rho_{M}r^{2}{\rm{d}}r\rightarrow+\infty.
\end{align}
Surprisingly, it is divergent, which is different from regular magnetic BH in rational NLED where $m_{M}$ is convergent \cite{Kruglov:2020tes}. Therefore, the magnetic mass cannot be applied mechanically on magnetically charged BI BH.

We then compare the metric (Eq. (\ref{fr})) with the metric of regular magnetic BH in rational NLED (Eq. (5) in \cite{Kruglov:2020tes}). To avoid confusion, we denote the latter as $\Phi(x)$,
\begin{align}
    \Phi(x)=1-\frac{P^{3/2}}{4x\sqrt{2\beta_{r}}}\left[\ln\frac{x^2-\sqrt{2}x+1}{x^2+\sqrt{2}x+1}+2\arctan(\sqrt{2}x+1)-2\arctan(1-\sqrt{2}x)\right],
    \label{rationalMetric}
\end{align}
where $\beta_{r}$ represents the effect of rational NLED, and $x=r/\sqrt[4]{\beta_{r}P^{2}}$. Expanding $f(r)$ and $\Phi(x)$ at $r\to+\infty$ and $r=0$, we find that both a magnetically charged BI BH and a regular magnetic BH in rational NLED degenerate into RN solutions at infinity. And the singularity is absent for regular magnetic BH in rational NLED \cite{Kruglov:2020tes}.

We summarize the comparisons above in Table \ref{comparison}.

\begin{table}[htbp]
    \setlength{\tabcolsep}{3mm}
    \begin{center}
        \begin{tabular}[b]{c|c|c}
            \midrule[2pt]
            Item & Rational NLED & BI \\
            \toprule[1.4pt]
            Condition of degenerating to MED & $\beta\mathcal{F}\to 0$ & $\beta\to+\infty$ \\
            \toprule[0.2pt]
            Magnetic mass $m_{\text{M}}$ & Convergent & Divergent \\
            \toprule[0.2pt]
            Metric function at $r=0$ & Converges to 1 & Divergent \\
            \toprule[0.2pt]
            Metric function at $r\to+\infty$ & Degenerates to RN & Degenerates to RN \\
            \midrule[2pt]
        \end{tabular}
    \end{center}
    \centering
    \caption{Summary of the comparisons between a regular magnetic BH in rational NLED and a magnetically charged BI BH.}
    \label{comparison}
\end{table}

\begin{figure}[htb]
    \begin{center}
        \subfigure[$\ P=0.6.$]{\includegraphics[width=7cm]{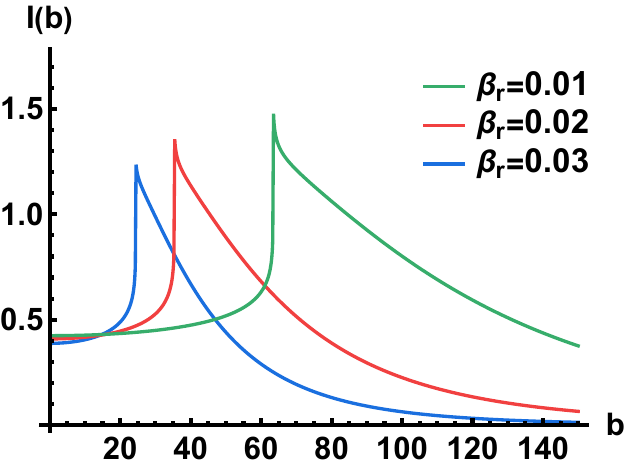}\label{I(b)_rational_change_beta}}
        \hspace{6pt}
        \subfigure[$\ \beta_{r}=0.01.$]{\includegraphics[width=7cm]{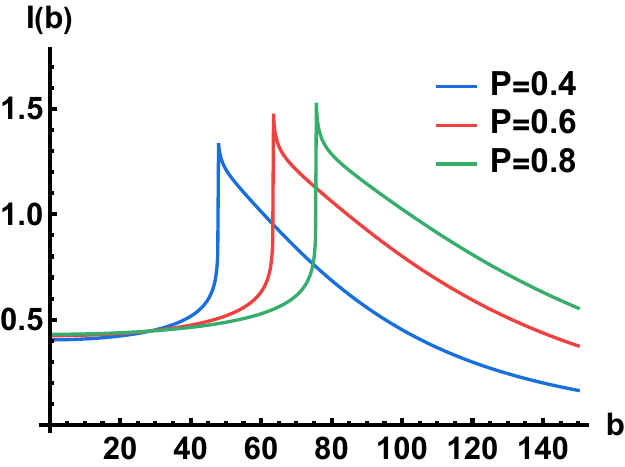}\label{I(b)_rational_change_P}}
    \end{center}
    \caption{The specific intensity $I(b)$ versus the impact parameter $b$ with different magnetic charge and rational NLED's effects.}
    \label{I(b)_rational}
\end{figure}

Next, we compare the observational appearances of the two models with static spherical accretions. According to \cite{Kruglov:2020tes}, the effective metric of photons takes the form as
\begin{align}
    {\rm{d}}s_{\text{eff}}^{2}=u(r)\left(-\Phi(r){\rm{d}}t^{2}+\frac{1}{\Phi(r)}{\rm{d}}r^{2}\right)+v(r)r^{2}\left({\rm{d}}\theta^{2}+\sin^{2}\theta{\rm{d}}\phi^{2}\right),
    \label{rationalEffMetric}
\end{align}
where $\Phi(r)$ is given in Eq. (\ref{rationalMetric}), and 
\begin{align}
    u(x)=\frac{x^{8}}{(x^{4}+1)^{2}},\quad v(x)=\frac{x^{8}(x^{4}-3)}{(x^{4}+1)^{3}},\quad x=\frac{r}{\sqrt[4]{\beta_{r}P^{2}}}.
\end{align}
Meanwhile the null geodesics in the equatorial plane is given by (having defined $b=L/E$ and reset the affine parameter $\lambda$ to $\lambda/L$)
\begin{align}
    \left(\frac{{\rm{d}}r}{{\rm{d}}\phi}\right)^{2}=r^{4}\left(\frac{v^{2}(r)}{b^{2}u^{2}(r)}-\frac{\Phi(r)v(r)}{u(r)r^{2}}\right).
    \label{rationalNullGeodesics}
\end{align}
From Eq. (\ref{rationalEffMetric}), the infinitesimal proper length is obtained by
\begin{align}
    {\rm{d}}l_{\text{prop}}=\pm \sqrt{G_{ij}{\rm{d}}x^{i}{\rm{d}}x^{j}}=\pm \sqrt{\frac{u(r)}{\Phi(r)}+v(r)r^{2}\left(\frac{{\rm{d}}\phi}{{\rm{d}}r}\right)^{2}}{\rm{d}}r.
    \label{rationaldl}
\end{align}
Through the same process as in deriving Eq. (\ref{g_static_derived}), the redshift factor of regular magnetic BH in rational NLED is given by
\begin{align}
    g=\sqrt{u(r)\Phi(r)}.
    \label{g_rational}
\end{align}
We apply lognormal distribution to photons' distribution again. Combining Eqs. (\ref{j_int}), (\ref{rationaldl}), and (\ref{g_rational}), integrate the specific intensity along the geodesics like Eq. (\ref{Ib}), the specific intensity $I(b)$ received by an observer at infinity with respect to the impact parameter $b$ for different magnetic charges $P$ and rational NLED's effects $\beta_{r}$ are plotted in Fig. \ref{I(b)_rational}.
\begin{figure}[htb]
    \begin{center}
        \subfigure[$\ \beta_{r}=0.01,\ P=0.4.$]{\includegraphics[width=.31\linewidth]{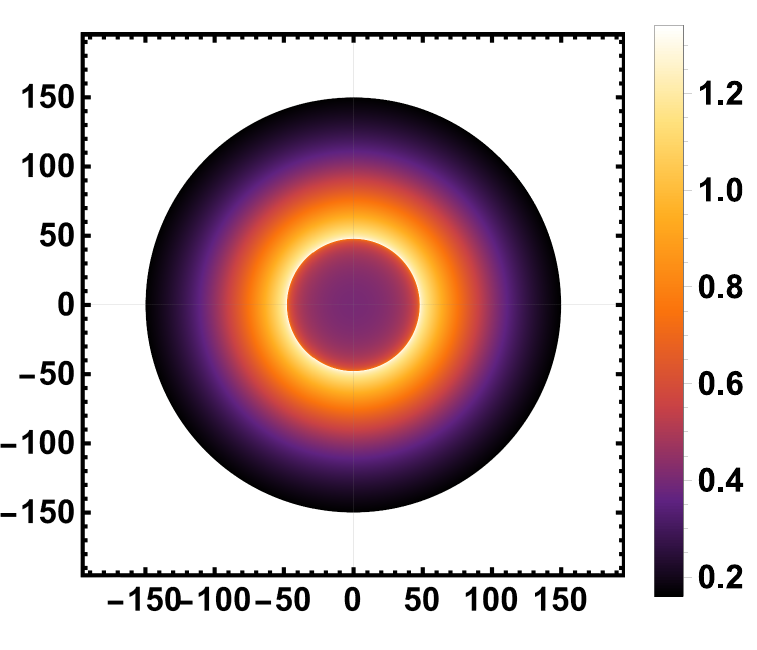}\label{shadow_rational_beta0.01P0.4}}
        \subfigure[$\ \beta_{r}=0.01,\ P=0.6.$]{\includegraphics[width=.31\linewidth]{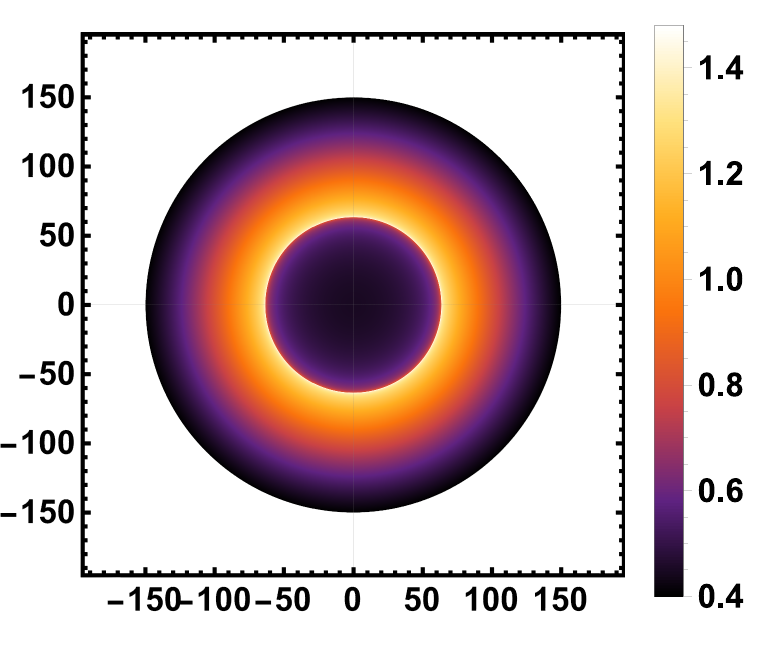}\label{shadow_rational_beta0.01P0.6}}
        \subfigure[$\ \beta_{r}=0.02,\ P=0.6.$]{\includegraphics[width=.3173\linewidth]{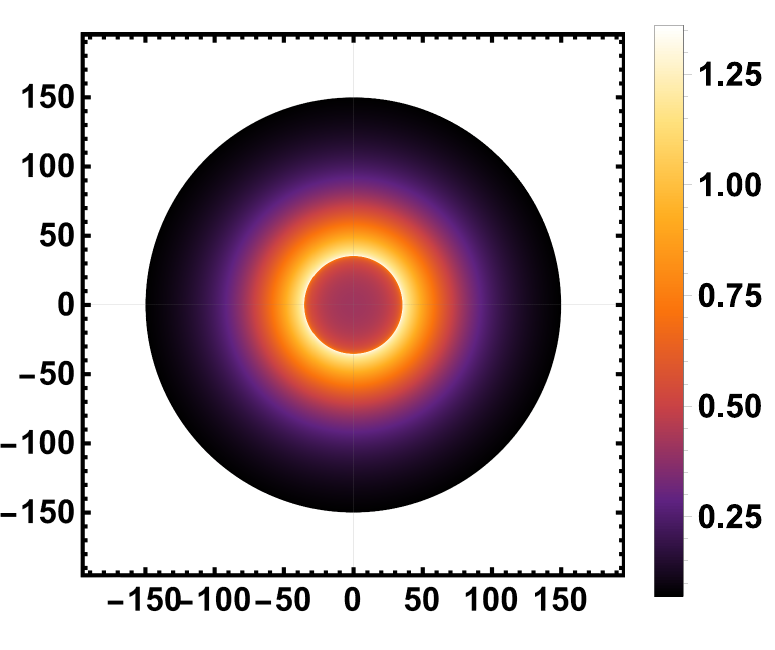}\label{shadow_rational_beta0.02P0.6}}
    \end{center}
    \caption{Observational appearances of regular BH in rational NLED with different magnetic charge and rational NLED's effects.}
    \label{shadows_rational}
\end{figure}

Comparing the tendency of the specific intensity of regular magnetic BH in rational NLED (Fig. \ref{I(b)_rational}) to that of magnetically charged BI BH (Figs. \ref{I(b)_static_change_beta} and \ref{I(b)_static_change_P}), two conclusions can be verified. Firstly, the peak of the specific intensity always appears at $b=b_{\text{ph}}$, for the photons rotate the BH infinite times if they incident from $b=b_{\text{ph}}$. Secondly, $b_{\text{ph}}$ is an intrinsic property of the spacetime that is only dependent on the magnetic charge and the effect of NLED. As the tendency of the specific intensity of the two models is similar, the observational appearances of the two models are also similar. It is conceivable that changing the NLED model will not have a great impact on the observational appearances of magnetic black holes with static spherical accretions. This point is also clarified by comparing Figs. \ref{shadows_static_change_beta}, \ref{shadows_static_change_P}, and \ref{shadows_rational}.
\end{appendices}

\bibliographystyle{unsrt}
\bibliography{egbib}

\end{document}